\documentclass[iop]{emulateapj}

\def\lapp{\ifmmode\stackrel{<}{_{\sim}}\else$\stackrel{<}{_{\sim}}$\fi}
\def\gapp{\ifmmode\stackrel{>}{_{\sim}}\else$\stackrel{>}{_{\sim}}$\fi}
\usepackage{multirow}
\usepackage{color}
\usepackage{amsmath}

\begin{document}

\title{NUSTAR OBSERVATIONS OF X-RAY BURSTS FROM THE MAGNETAR 1E~1048.1$-$5937}

\author{
Hongjun An\altaffilmark{1}, Victoria M. Kaspi\altaffilmark{1,10},
Andrei M. Beloborodov\altaffilmark{2}, Chryssa Kouveliotou\altaffilmark{3},
Robert F. Archibald\altaffilmark{1}
Steven E. Boggs\altaffilmark{4}, Finn E. Christensen\altaffilmark{5},
William~W.~Craig\altaffilmark{4,6}, Eric~V.~Gotthelf\altaffilmark{2},
Brian W. Grefenstette\altaffilmark{7},
Charles J. Hailey\altaffilmark{2}, Fiona A. Harrison\altaffilmark{7},
Kristin K. Madsen\altaffilmark{7}, Kaya Mori\altaffilmark{2},
Daniel Stern\altaffilmark{8}, and William~W.~Zhang\altaffilmark{9}
}
\affil{
{\small $^1$Department of Physics, McGill University, Montreal, Quebec, H3A 2T8, Canada}\\
{\small $^2$Columbia Astrophysics Laboratory, Columbia University, New York NY 10027, USA}\\
{\small $^{3}$Space Science Office, ZP12, NASA Marshall Space Flight Center, Huntsville, AL 35812, USA}\\
{\small $^4$Space Sciences Laboratory, University of California, Berkeley, CA 94720, USA}\\
{\small $^5$DTU Space, National Space Institute, Technical University of Denmark, Elektrovej 327, DK-2800 Lyngby, Denmark}\\
{\small $^6$Lawrence Livermore National Laboratory, Livermore, CA 94550, USA}\\
{\small $^7$Cahill Center for Astronomy and Astrophysics, California Institute of Technology, Pasadena, CA 91125, USA}\\
{\small $^{8}$Jet Propulsion Laboratory, California Institute of Technology, Pasadena, CA 91109, USA}\\
{\small $^{9}$Goddard Space Flight Center, Greenbelt, MD 20771, USA}\\
}
\altaffiltext{10}{Lorne Trottier Chair; Canada Research Chair}

\begin{abstract}
We report the detection of eight bright X-ray bursts from the 6.5-s magnetar 1E~1048.1$-$5937,
during a 2013 July observation campaign with
{\em the Nuclear Spectroscopic Telescope Array (NuSTAR)}.
We study the morphological and spectral properties of these bursts and their evolution with time.
The bursts resulted in count rate increases by orders of magnitude, sometimes limited
by the detector dead time, and showed blackbody spectra with $kT\sim$ 6--8 keV in the $T_{\rm 90}$
duration of 1--4 s, similar to earlier bursts detected from the source.
We find that the spectra during the tail of the bursts can be
modeled with an absorbed blackbody with temperature decreasing with flux. The bursts flux decays
followed a power-law of index ~0.8--0.9. In the burst tail spectra, we detect a $\sim$13 keV emission
feature, similar to those reported in previous bursts from this source as well as from
other magnetars observed with the {\em Rossi X-ray Timing Explorer (RXTE)}.
We explore possible origins of the spectral feature such as proton cyclotron emission, which implies
a magnetic field strength of $B\sim2\times10^{15}\ \rm G$ in the emission region.
However, the consistency of the energy of the feature in different objects requires further explanation.
\end{abstract}


\section{Introduction}
Magnetars are isolated neutron stars which have very high
magnetic-field strengths inferred from the high spin down rate,
typically greater than $10^{14}\ \rm G$ \citep[e.g.,][for a catalog of magnetars]{vg97, kds+98, ok14}.
Their X-ray luminosities are often greater than their spin-down power,
and theorized, therefore, to be powered
by the decay of intense internal magnetic fields \citep{dt92, td95, td96}.
The decay of the magnetic field may gradually build up stress in the crust, which can fracture
it and/or twist the external magnetic field,
causing short X-ray and soft-gamma-ray bursts \citep[][]{fhd+01,lwg+03,gwk+11}
and sudden flux increases \citep[][]{tlk02, b09}.
Historically, two classes of X-ray pulsars were thought to be magnetars:
Anomalous X-ray Pulsars (AXP) whose X-ray luminosities exceed the spin-down power,
and Soft Gamma Repeaters (SGR) which show repeated soft gamma-ray bursts.
However, distinction between the two classes has been significantly blurred \citep[][]{gkw02,kgw+03}.
There are 26 magnetars discovered to date with various spectral and temporal properties
\citep[see][]{ok14}.\footnote{See the online magnetar catalog for a compilation
of known magnetar properties:\\ http://www.physics.mcgill.ca/$\sim$pulsar/magnetar/main.html.}

Magnetar bursts have a  variety of morphologies, including short ($\sim$100 ms) symmetric bursts,
multiple peaked bursts, those with fast rises and longer decays, and
some which exhibit very long extended `tails' \citep[see e.g.,][]{wkg+05, gkw06}.
Previous studies have suggested relationships
between burst intensity and tail energetics \citep[][]{lwg+03, wkg+05}
and even the possibility of two distinct types of bursts \citep[][]{wkg+05}. 
Burst spectra are generally described with thermal models \citep[][]{vkg+12} although there has been
evidence for spectral features in some bursts \citep[][]{gkw02,wkg+05,dkg09,gdk11}. 

The magnetar 1E~1048.1$-$5937 is relatively active, often showing X-ray bursts and
unstable timing behavior. Its spin period is 6.46 s, and the spin inferred surface
magnetic-field strength is $B=4\times 10^{14}\ \rm G$. In quiescence, it shows a spectrum which is
well described with a blackbody plus power-law model having
$kT\sim 0.6\ \rm keV$ and $\Gamma \sim 2.9$ \citep[][]{tgd+08}.
The distance to the source is estimated to
be 9 kpc \citep[][]{dv06}, which we use throughout this paper.
Interestingly, it is the first AXP in which
X-ray bursts were seen \citep{gkw02}, and has shown several more bursts since then \citep[][]{dkg09},
hereby blurring the distinction
between the AXP and SGR classes \citep[see also][for 1E~2259$+$586]{kgw+03}.
Another interesting property of 1E~1048.1$-$5937's bursts is a possible emission feature at $\sim$13 keV
in the spectrum which was previously seen during its 2002 burst in {\em RXTE} data \citep[][]{gkw02,dkg09}.
Similar spectral features have been seen in X-ray bursts from other magnetars
as well, all with {\em RXTE} \citep[XTE~J1810$-$197, 4U~0142$+$61;][]{wkg+05, gdk11}.

In this paper, we report on the spectral and temporal properties of eight new bursts
from 1E~1048.1$-$5937 detected serendipitously with
{\em the Nuclear Spectroscopic Telescope Array (NuSTAR)} in 2013 July.
We describe the observations and
data reduction in Section~\ref{sec:obs}, and show the data analysis and results in
Section~\ref{sec:ana}. We then discuss the implications of the analysis results
(Sec.~\ref{sec:disc}) and conclude in Section~\ref{sec:concl}.

\medskip
\section{Observations}
\label{sec:obs}
{\em NuSTAR} operates in the 3--79 keV band, and is the most sensitive satellite
in the $\sim$10--79 keV band thanks to its unique
focusing capability in this energy range. The energy resolution is 400 eV at 10 keV (FWHM),
and the temporal resolution
is 2 $\mu$s \citep[see ][for more details]{hcc+13}. Its excellent temporal and spectral
resolutions together with the high sensitivity above $\sim$10 keV are optimal for the study of
burst properties of magnetars because such events are
very brief ($\sim$ms) and sometimes spectrally very hard.

1E~1048.1$-$5937 was observed with {\em NuSTAR} on 2013 July 17--27 with a total net exposure
of $\sim$320 ks (Obs. ID 30001024001--7). A 70-ks joint {\em XMM-Newton} observation (Obs. ID: 0723330101)
was conducted using the small window mode for MOS1/2 and the full frame mode for PN on 2013 July 22 to
extend the spectral coverage down to $\sim$0.5 keV where the thermal component is dominant.
The source was not known to be in a particularly active state at the time of the observation.
During the {\em NuSTAR} observations, eight X-ray bursts from the source were detected to our surprise,
with one simultaneous detection in the {\em XMM-Newton} data.

The {\em NuSTAR} data were processed with {\ttfamily nupipeline} 1.3.1 along with CALDB version 20131007
using standard filters except for {\tt PSDCAL}.
We set {\tt PSDCAL=NO} in order to recover more exposure by
slightly sacrificing the pointing accuracy.\footnote{http://heasarc.gsfc.nasa.gov/docs/nustar/analysis/nustar\_sw\\guide.pdf}
We verified that the exposure increases and that the imaging,
timing and spectral analysis results are consistent with those obtained with {\tt PSDCAL=YES}.
The {\em XMM-Newton} data were processed with Science Analysis System (SAS) 12.0.1 using
the standard filtering process. We then further processed the event files for analysis as described below.

\medskip

\section{Data Analysis and Results}
\label{sec:ana}

\subsection{Burst Morphology}
\label{burstmorphology}

\begin{figure}[ht]
\hspace{0.0 mm}
\begin{tabular}{c}
\vspace{-5.0 mm}
\includegraphics[width=2.5 in, angle=90]{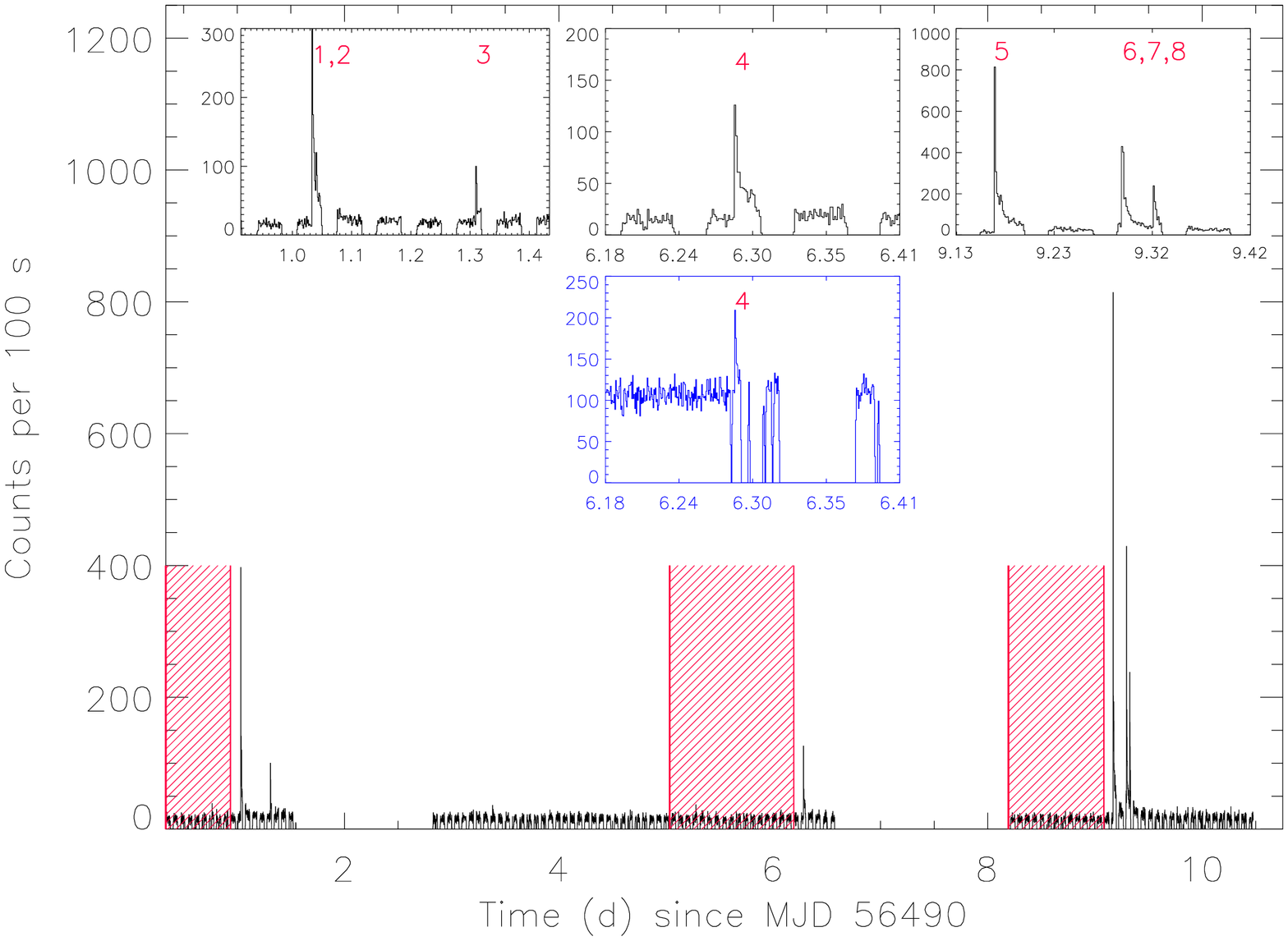} \\
\hspace{-2.0 mm}
\includegraphics[width=2.6 in, angle=90]{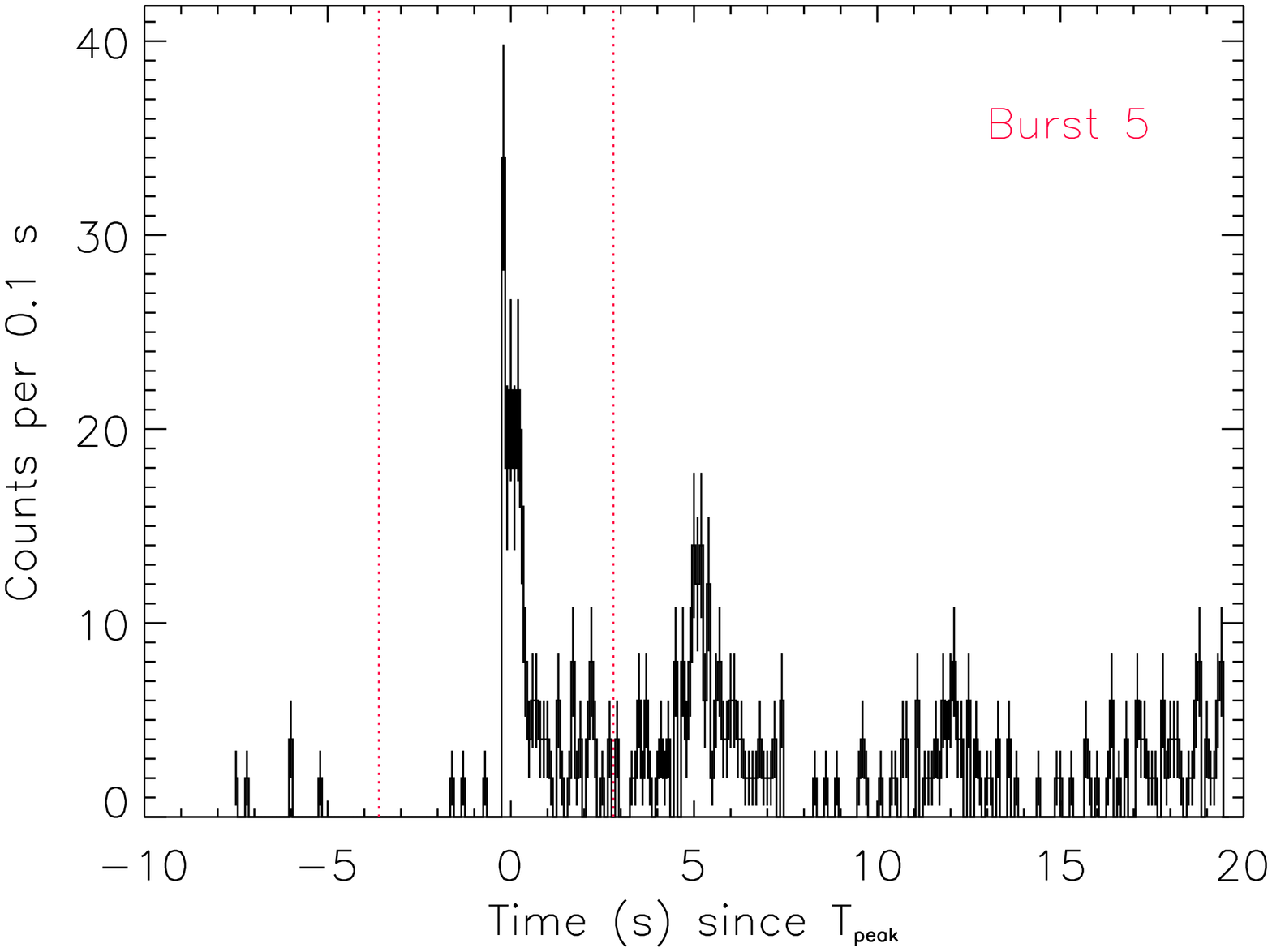} \\
\end{tabular}
\vspace{-2.0 mm}
\figcaption{{\it Top:} 3--79 keV light curve for the {\em NuSTAR} observations with Modules A and B combined.
Time intervals for pre-burst background extraction are shown in red
hatched lines (see Sections~\ref{burstspec} and \ref{specevol}).
Insets show zoomed-in 3--79 keV {\em NuSTAR} and 0.5--10 keV {\em XMM-Newton} light curves (blue)
around the bursts with burst numbers. {\it Bottom:} A 30-s light curve around burst 5
in the 3--79 keV band. $T=0$ is set to the burst peak time, and red vertical lines show a one pulse period
in which we characterize the burst morphology (see Eq.~\ref{eqn:lightcurve} and Table~\ref{ta:bursts}).
\label{fig:lcall}
}
\vspace{0.0 mm}
\end{figure}

In order to search for bursts, we extracted events in an aperture of 60$''$ around the source position
in the {\em NuSTAR} image, and produced barycenter-corrected light curves with a bin size of 0.5 s.
We then searched for time bins having a significantly larger number of counts compared with
the persistent level which was extracted from the source region (and which is dominated by the persistent
flux from the source) in 10 pulse periods ($\sim$64 s) prior
to the time bin that is being searched. The average count rate of the persistent emission was
0.2 cps in the 3--79 keV band within the $R=60''$ aperture (see Fig.~\ref{fig:lcall}).
We further verified that the count rate in the off-source region did not show any significant
increase (e.g., due to a background flare) over any short time interval. We then calculated the Poisson
probability of the observed count given the background rate for each time bin.
To be considered significant, we required a $>$3$\sigma$ chance probability $p$ after
considering the number of trials. In total, we detected eight bursts with high significance,
each one having $p<10^{-10}$.
We also tried different bin sizes (e.g., 0.01--100 s), and found similar results,
although the significance changes depending on the binning, and some bursts are not significantly detected
on some timescales. The bursts 2 and 7 are not detected on a long timescale (100 s), and we do not
report their tail properties.

\newcommand{\markaa}{\tablenotemark{a}}
\newcommand{\markbb}{\tablenotemark{b}}
\newcommand{\markcc}{\tablenotemark{c}}
\newcommand{\markdd}{\tablenotemark{d}}
\begin{table*}
\vspace{0.0 mm}
\begin{center}
\caption{Summary of {\em NuSTAR}-detected bursts
\label{ta:bursts}}
\scriptsize{
\begin{tabular}{ccccccccccccccc} \hline\hline
Burst & $T_{\rm 0}$\markaa & $\phi$\markbb & $T_{\rm r}$\markaa & $T_{\rm f}$\markaa & $A$\markaa & $C_{\rm 1}$ & $C_{\rm 2}$\markaa & $T_{\rm 90}$\markcc & $kT$\markdd & $L_{\rm 90}$\markdd  \\
	& (day)&  & (s)& (s)	& (cts$\ \rm s^{-1}$) & (cps) & (cps) & (s) & (keV) & ($10^{37}\ \rm erg\ s^{-1}$) \\ \hline
1 & 1.0330396 & $0.477^{+0.006}_{-0.009}$ & $1.66^{+0.2}_{-0.2}$ & $0.07^{+0.19}_{-0.05}$ & $56^{+6}_{-6}$ & $1.8^{+0.7}_{-0.6}$ & $8^{+2}_{-2}$ & $4.0^{+0.5}_{-0.5}$  & 6.3(8) & 1.6(4)\\ 
2& 1.0405240 & $0.555^{+0.001}_{-0.001}$ &  $0.012^{+0.010}_{-0.007}$ & $0.03^{+0.01}_{-0.01}$ & $170^{+70}_{-60}$ & $2.7^{+0.8}_{-0.7}$ & $3^{+2}_{-1}$  & $0.10^{+0.03}_{-0.03}$ & $\cdots$ & $\cdots$\\ 
3 & 1.3100872 & $0.9177^{+0.001}_{-0.0006}$ & $0.06^{+0.02}_{-0.02}$ & $<$0.018 & $190^{+60}_{-50}$ & $1.6^{+0.5}_{-0.4}$ & $4^{+1}_{-1}$ & 0.17 & $\cdots$ & $\cdots$  \\ 
4 & 6.2812256 & $0.6631^{+0.0012}_{-0.0006}$ & $0.10^{+0.02}_{-0.02}$ & $<$0.019 & $170^{+50}_{-40}$ & $0.7^{+0.3}_{-0.2}$ & $5^{+2}_{-1}$ & 0.26 & $\cdots$ & $\cdots$ \\ 
5 & 9.1684044 & $0.5052^{+0.0002}_{-0.0005}$ & $<$0.007  & $0.42^{+0.04}_{-0.03}$ & $360^{+30}_{-30}$ & $3^{+2}_{-1}$ & $17^{+4}_{-3}$  & 0.98 & 8.0(8) & 16(3) \\ 
6 & 9.2942137 & $0.7269^{+0.0034}_{-0.0002}$ & $0.8^{+0.1}_{-0.1}$ & $<$0.055 & $100^{+10}_{-10}$ & $8^{+2}_{-2}$ & $16^{+4}_{-4}$  & 2.0 & 7(1) & 4(1)   \\ 
7& 9.2973844 & $0.155^{+0.005}_{-0.002}$ & $<0.03$ & $0.17^{+0.05}_{-0.04}$ & $110^{+30}_{-30}$ & $3^{+1}_{-1}$ & $7^{+1}_{-1}$  & 0.46 & $\cdots$ & $\cdots$  \\ 
8 & 9.3254520 & $0.4116^{+0.0001}_{-0.0001}$ & $<$0.015  & $0.65^{+0.06}_{-0.05}$ & $170^{+20}_{-20}$ & $1.5^{+0.8}_{-0.6}$ & $4^{+2}_{-2}$  & 1.52 & 8(1) & 5(1) \\ \hline  
\end{tabular}}
\end{center}
\vspace{-1.0 mm}
\footnotesize{{\bf Notes.} 
Uncertainties are at the 1$\sigma$ confidence level, and upper limits are at the 90\% confidence level.}\\
$^{\rm a}${ Parameters for the short timescale light curve as defined in Equation~\ref{eqn:lightcurve}. $T_{\rm 0}$ is days since MJD 56490 (TDB).}\\
$^{\rm b}${ Spin phase corresponding to $T_{\rm 0}$, where phase zero is defined at the pulse minimum (56490.3343345727 MJD),
same as that for the timing analysis in Figures~\ref{fig:residual}--\ref{fig:xmmprofiles}.}\\
$^{\rm c}${ Time interval which includes 90\% of the burst counts (the exponential functions in Equation~\ref{eqn:lightcurve}.
$T_{\rm 90}$'s for the rising and the falling function were calculated separately and then summed to obtain that for the burst.
When only an upper limit is available for $T_{\rm r}$ or $T_{\rm f}$, we used the upper limit to calculate $T_{\rm 90}$ and show it
without uncertainties.}\\
$^{\rm d}${ Spectral parameters for a blackbody spectrum corresponding to $T_{\rm 90}$: blackbody temperature
$kT$ and bolometric luminosity $L_{\rm 90}$.}\\ 
\vspace{-3.0 mm}
\end{table*}

We note that many time bins immediately after a burst were detected significantly above quiescence.
The significantly detected time bins are likely those of the burst tail emission or the peaks of the
source pulsations which are visible immediately after the burst (e.g., see Fig.~\ref{fig:lcall} bottom).
However, there may be time bins with significantly
higher count rates which are produced by an independent burst event.
To investigate whether these significantly detected bins after a burst
were independent events and not related to the burst tail or
the pulse peaks, we proceeded as follows.
We extracted a $\sim$50~s light curve after
a burst and characterized it with a decay model,
specifically an exponentially decaying sine plus an exponential decay plus a constant.
We then searched for time bins ($\Delta T=$0.05--1~s) having counts above the
decay model with $\gapp$3$\sigma$ confidence, and found none. Therefore, we conclude that there
were only eight significant bursts during
the observation. The {\em NuSTAR} light curve is shown in Figure~\ref{fig:lcall},
and the burst properties are listed in Table~\ref{ta:bursts}. The burst light curves all exhibit
rises and decays of a few seconds with relatively long tails ($\sim$ks).

After we identified the bursts, we extracted events in a one period interval that contains the burst
in order to further characterize the temporal properties of the bursts on very short time scales.
Note that we included one more period for the bursts that occurred late in pulse phase so as
not to miss the falling tail.
We fit the time series to an exponentially rising and falling function
\begin{equation}
\label{eqn:lightcurve}
F(t) =
 \begin{cases}
   A e^{(t-T_0)/T_{\rm r}} + C_1 &  t<T_0, \\
   (A - C_2) e^{-(t-T_0)/T_{\rm f}} + C_1 + C_2 & t\geq T_0,
 \end{cases}
\end{equation}
where $A$ is the amplitude, $T_{\rm 0}$ is the burst peak time,
$T_{\rm r}$ is the rising time, $T_{\rm f}$ is the falling time,
and $C_{1,2}$ are constants. We note that the decay of a burst has typically been modeled
with a double exponential function \citep[e.g.,][]{gdk11}, but here we
replace the second exponential with a constant ($C_2$) because
this suffices for describing our data in the chosen time span, which is
much smaller than the decay constant of any second exponential.
Since the time scales are very short, and there are only a few events in each $\sim$ms time bin,
we used maximum-likelihood optimization. Furthermore, we used events in the whole detector because
having a well sampled time series is important in the fitting.
We modeled the background with another constant, $C_1$.
The results of the fitting are presented in Table~\ref{ta:bursts}. 

We note that the observed count rate is smaller than the incident rate for the brightest burst
due to detector deadtime \citep[][]{hcc+13}.
Since the maximum count rate for burst 4 is comparable to the maximum count rate
that the {\em NuSTAR} detectors can process ($\sim$400 cts $\rm s^{-1}$ per module),
we consider the effect of deadtime in order to calculate the incident count rate,
$R_{\rm i}$, via the following relation:
$$R_{\rm i} = \frac{R_{\rm o}}{1-R_{\rm o} \tau_{\rm D}}.$$
Here $R_{\rm o}$ is the observed count rate, and $\tau_{\rm D}$ is the detector dead time ($\sim$2.5 ms)
for each observed event. The incident peak count rates are higher,
and the rising and the falling times are smaller than the observed values in Table~\ref{ta:bursts},
but within a factor of $\sim$2 of the true values.
For example, the maximum observed count rate for burst 4 is
$\sim$200 cts~$\rm s^{-1}$ per module, and the incident rate is estimated to be $\sim$400~cts~$\rm s^{-1}$ per
module for this burst.

\medskip
\subsection{Timing Analysis}
\label{timingana}

\begin{figure}
\centering
\vspace{-3.0 mm}
\begin{tabular}{c}
\hspace{-15.0 mm}
\vspace{-4.0 mm}
\includegraphics[width=3.7 in]{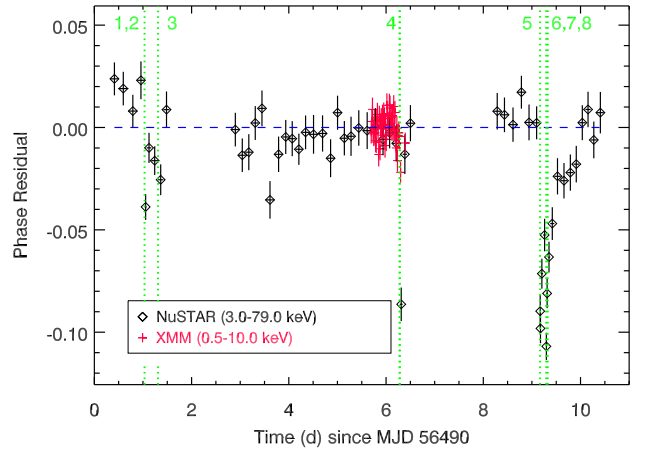} \\
\end{tabular}
\hspace{-15.0 mm}
\vspace{0.00 mm}
\figcaption{The measured timing residuals for $P=6.46168155\ \rm s$ for the {\em NuSTAR} data
in the 3--79 keV band (black) and the {\em XMM-Newton} data in the 0.5--10 keV band (red).
The best-fit function is shown in a blue horizontal dashed line and the burst times are shown
in green vertical dotted lines.
\label{fig:residual}
}
\end{figure}

\begin{figure*}
\centering
\begin{tabular}{ccc}
\hspace{-5.0 mm}
\includegraphics[width=2.1 in, angle=90]{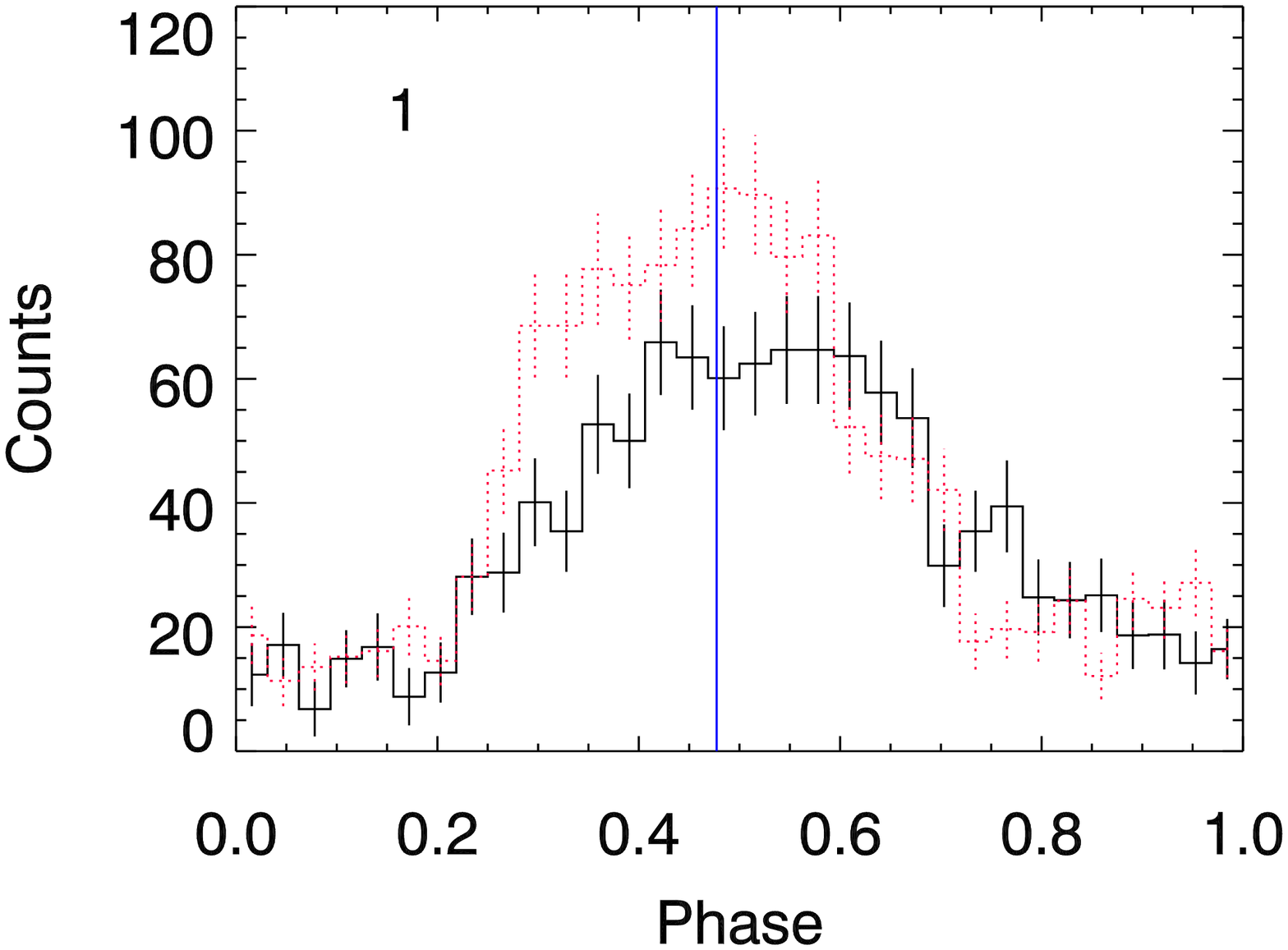} &
\hspace{-17.0 mm}
\includegraphics[width=2.1 in, angle=90]{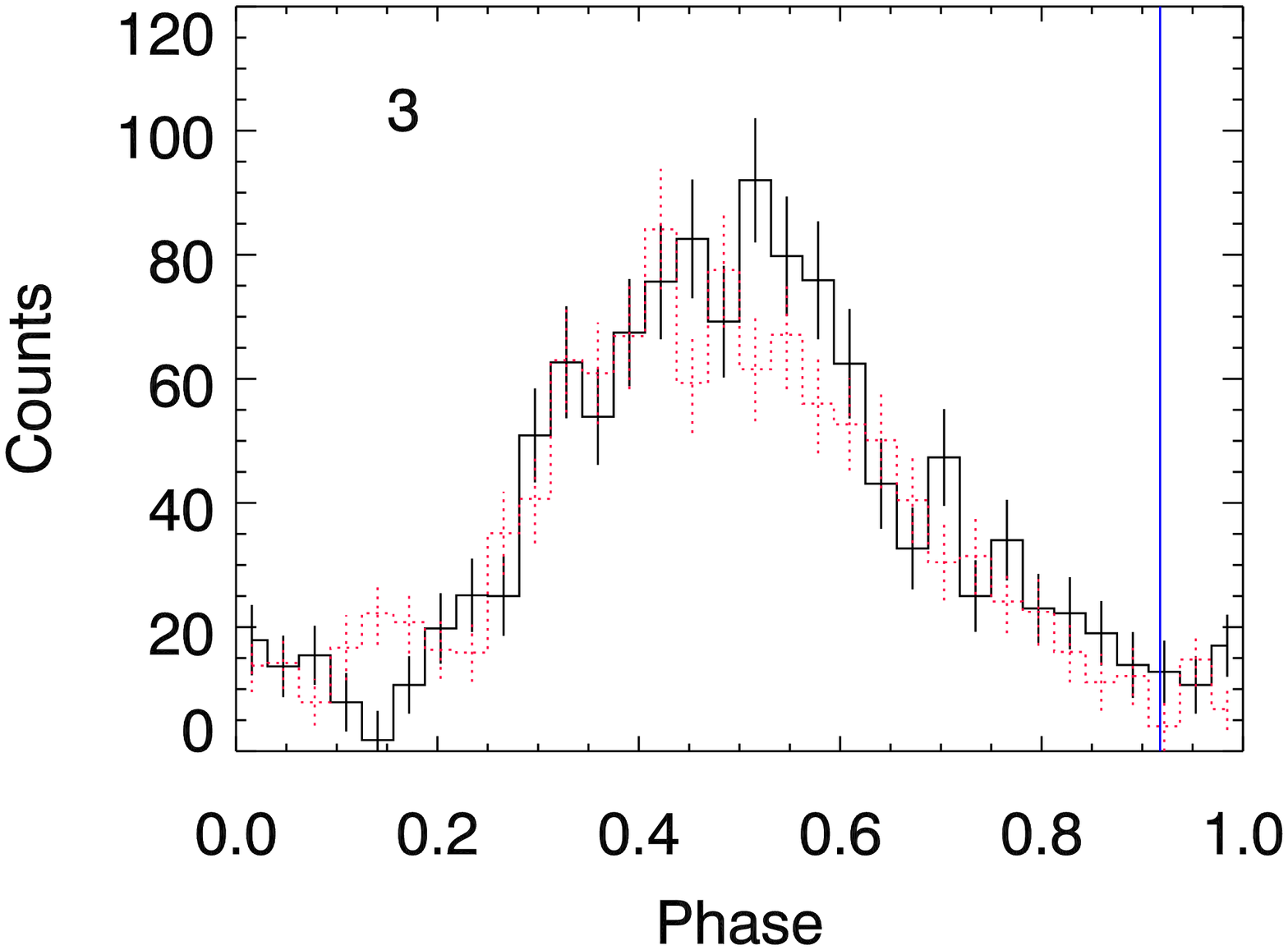} &
\hspace{-17.0 mm}
\vspace{-7.0 mm}
\includegraphics[width=2.1 in, angle=90]{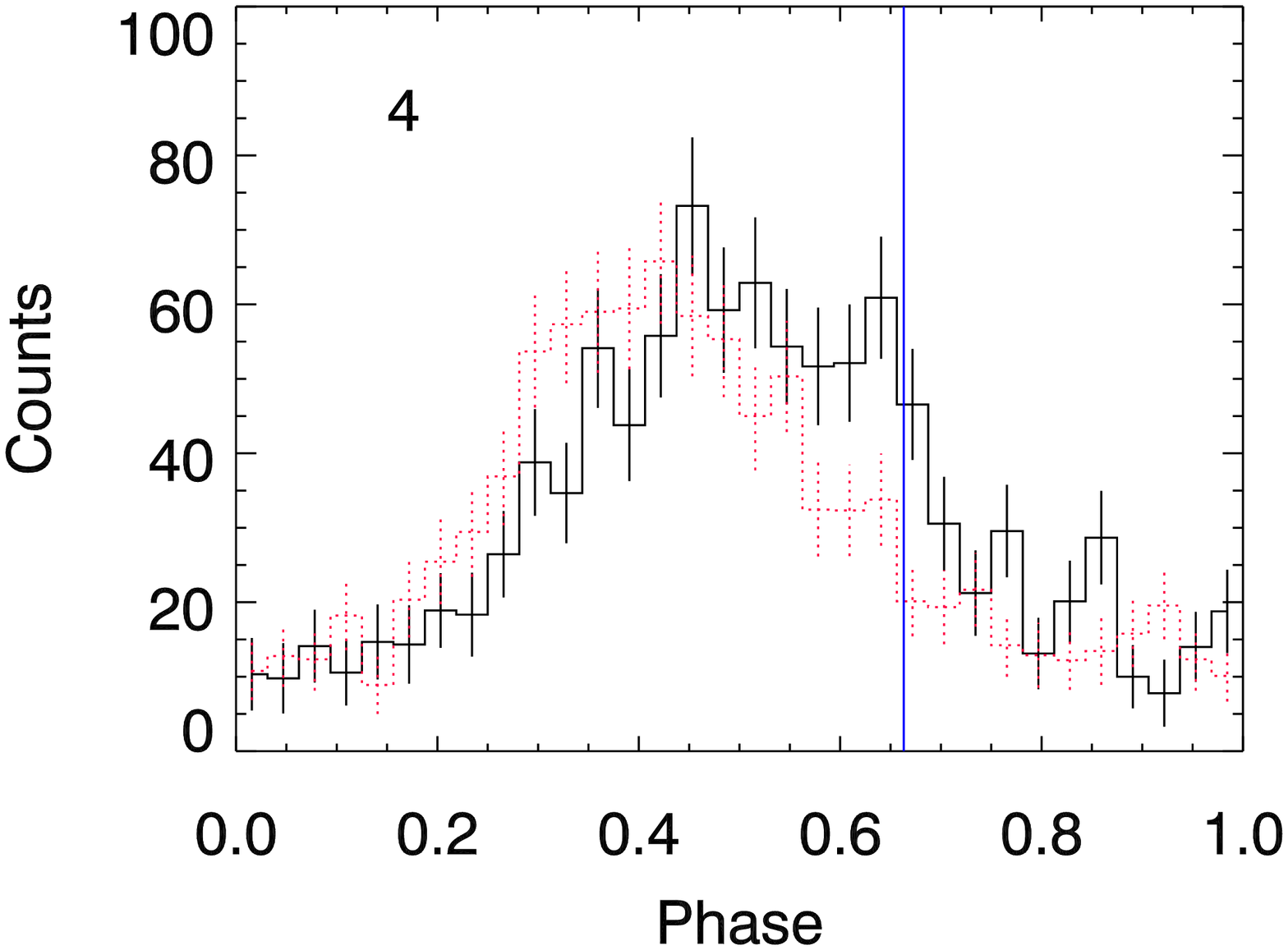} \\
\hspace{-5.0 mm}
\includegraphics[width=2.1 in, angle=90]{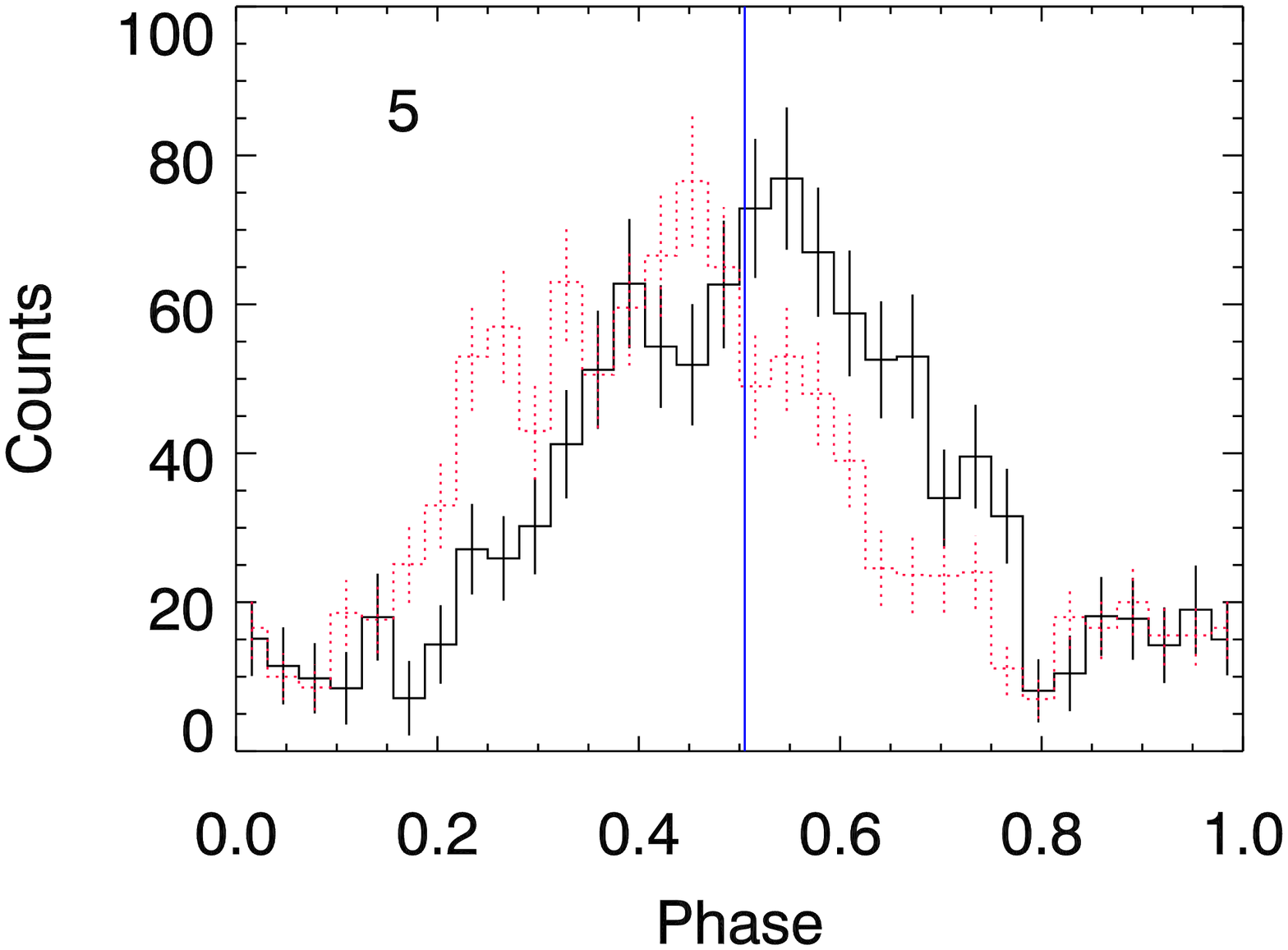} &
\hspace{-17.0 mm}
\includegraphics[width=2.1 in, angle=90]{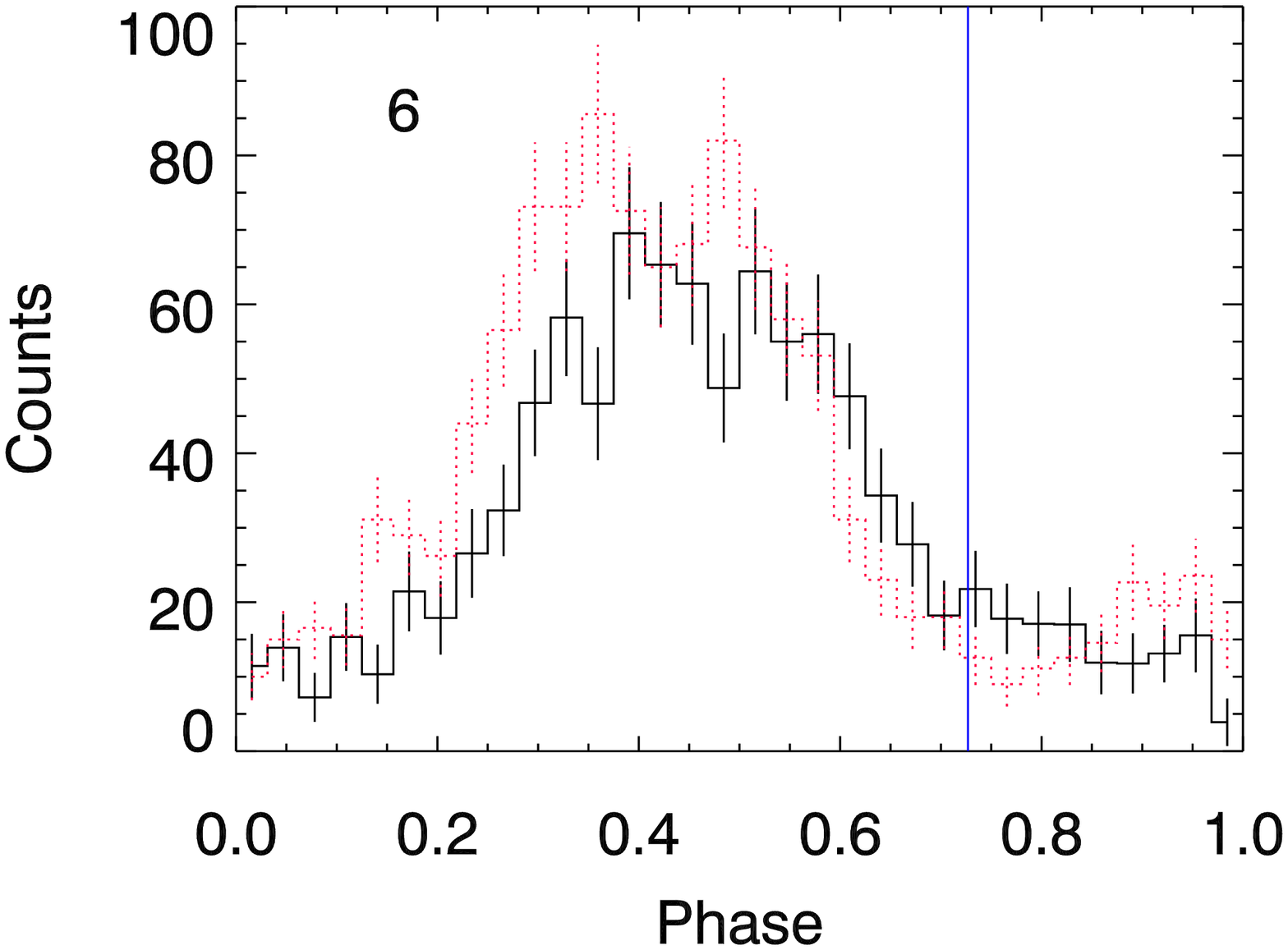} &
\hspace{-17.0 mm}
\vspace{-7.0 mm}
\includegraphics[width=2.1 in, angle=90]{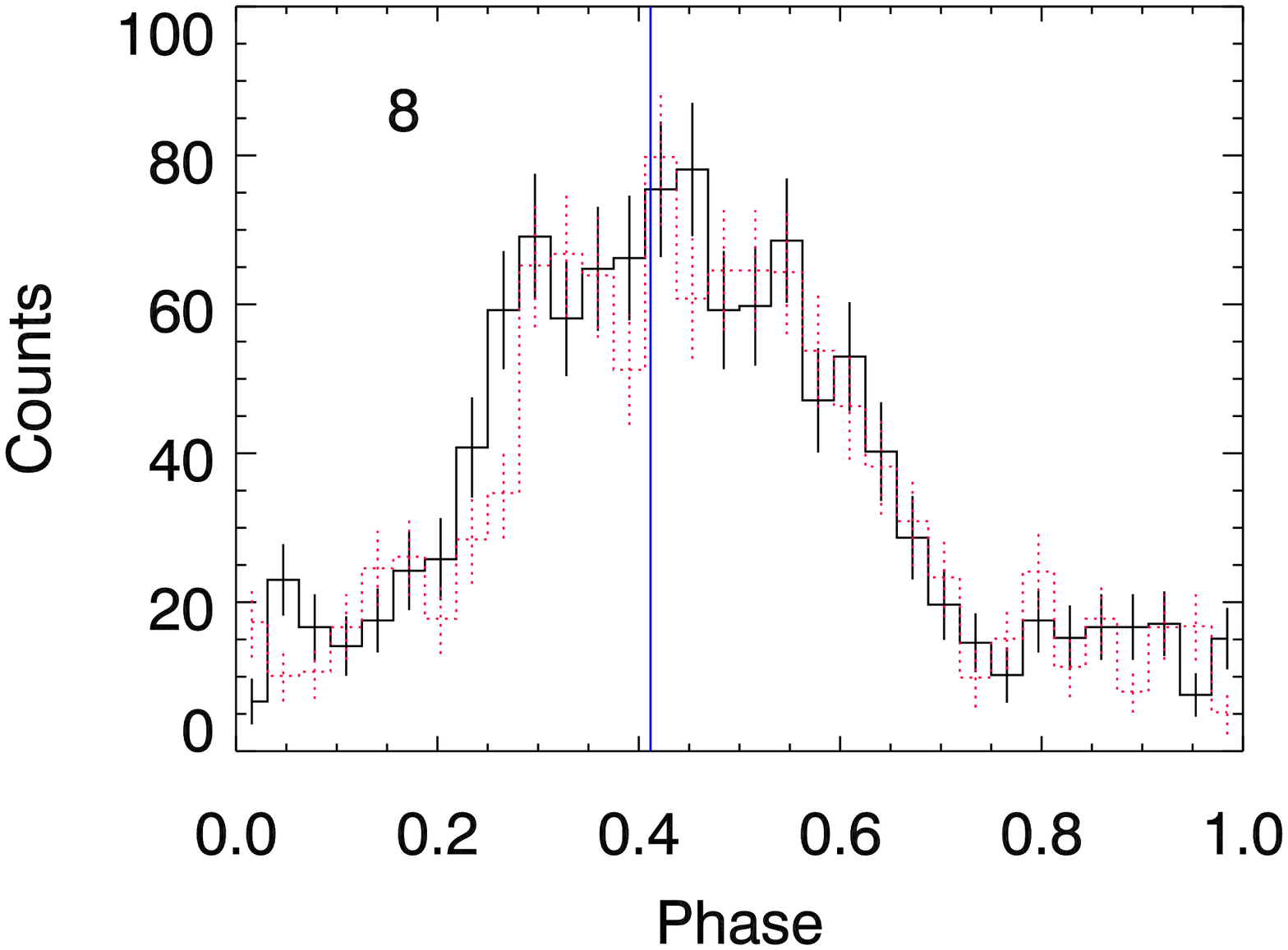} \\
\end{tabular}
\hspace{-5.0 mm}
\vspace{2.0 mm}
\figcaption{Immediate pre- (black) and post-burst (red) {\em NuSTAR} pulse profiles for bursts 1, 3, 4, 5, 6, and 8
in the 3--79 keV band.
$T_{\rm 90}$ burst intervals were not included in making these profiles.
Backgrounds were subtracted, and the integration times were chosen so that each pulse profile had
$\sim$1200 cts. Burst numbers are shown in the plots for reference.
Vertical lines in the plots are the phases corresponding to the $T_{\rm 0}$
in equation~\ref{eqn:lightcurve} for each burst. Note that the pre-burst profiles of burst 5 and 6
include significant tail emission of burst 4 and 5, respectively.
\label{fig:profiles}
}
\end{figure*}

\begin{figure*}
\centering
\vspace{-5.0 mm}
\begin{tabular}{ccc}
\hspace{-5.0 mm}
\includegraphics[width=2.1 in, angle=90]{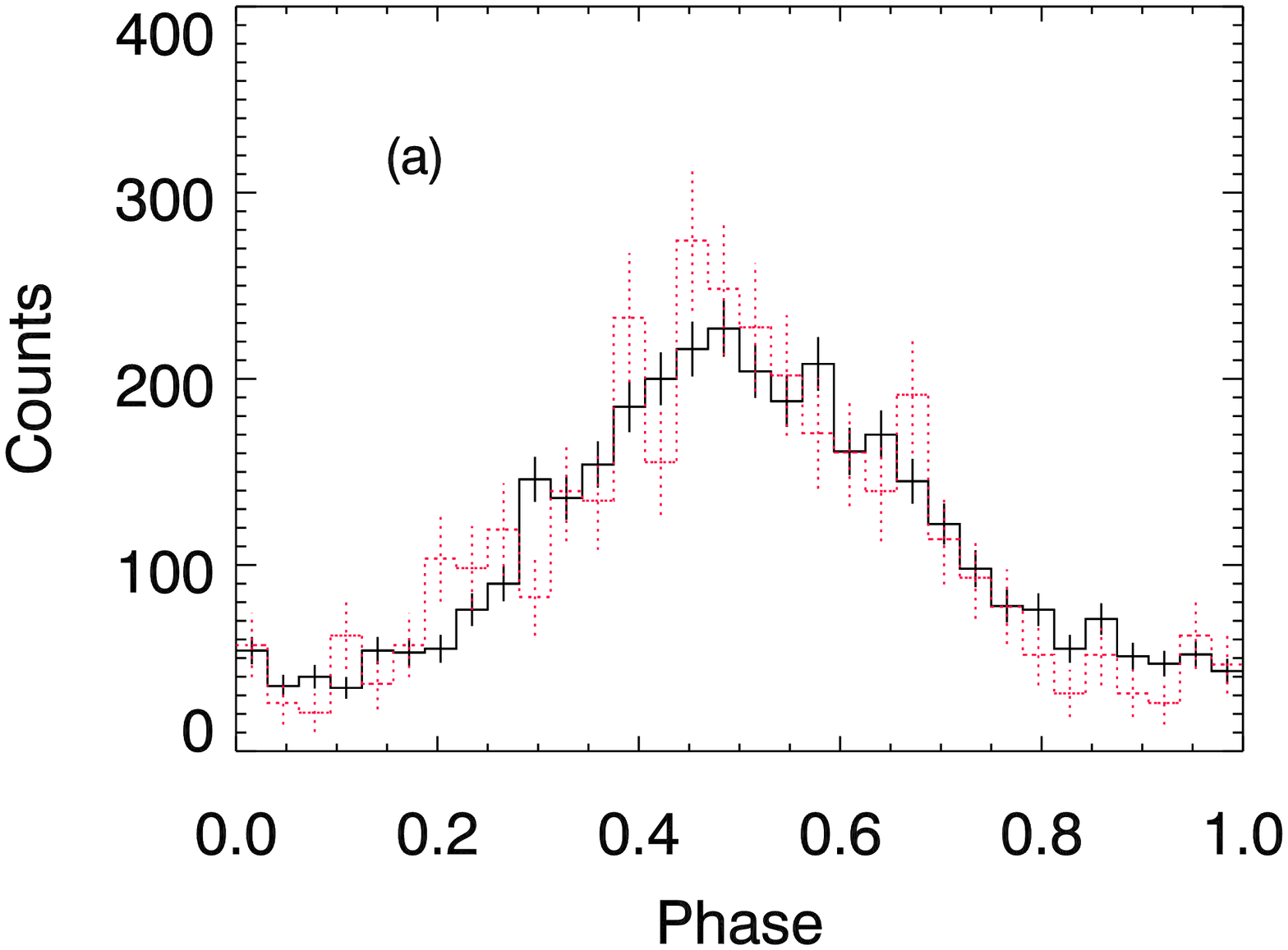} &
\hspace{-17.0 mm}
\includegraphics[width=2.1 in, angle=90]{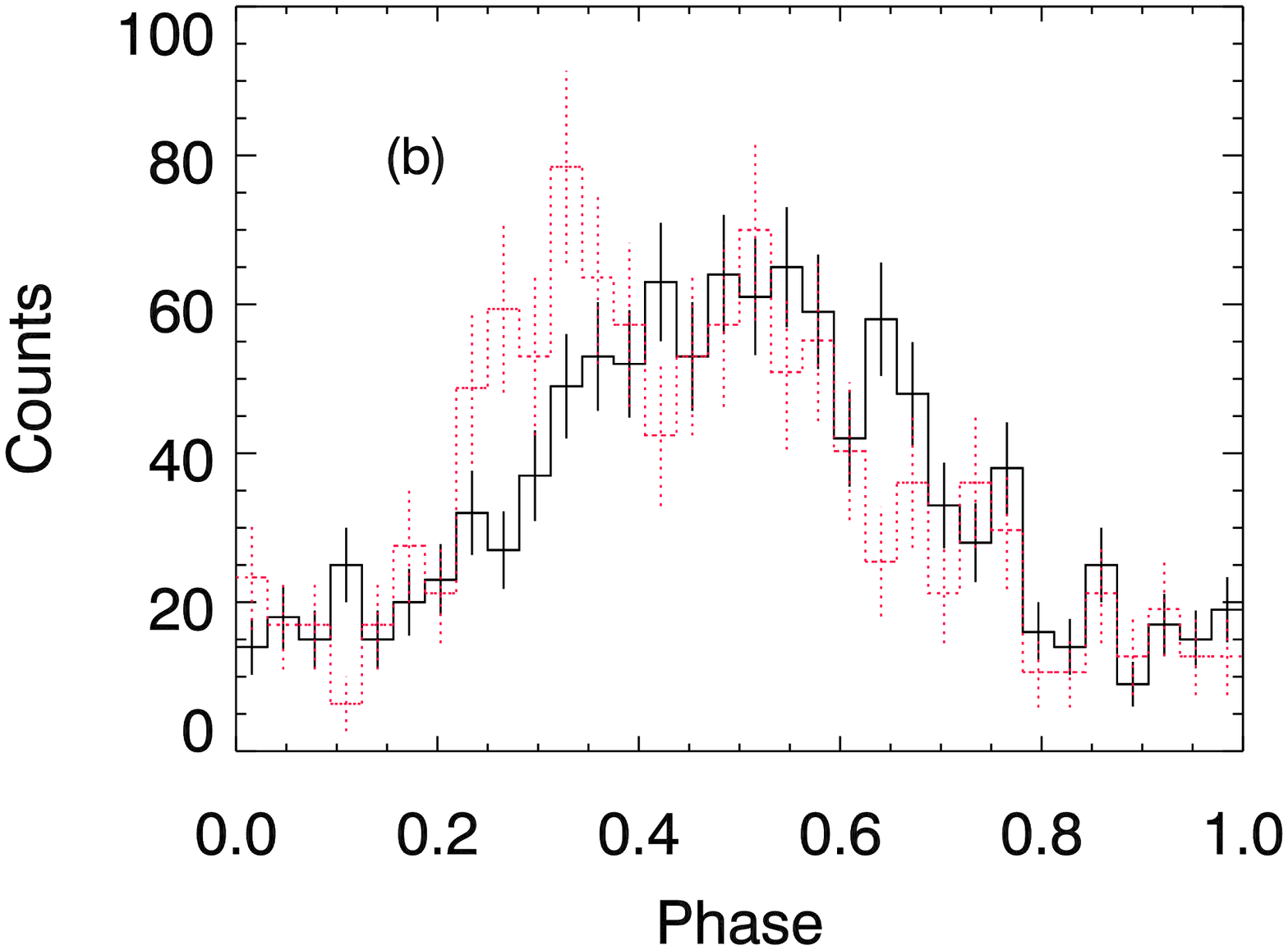} &
\hspace{-17.0 mm}
\includegraphics[width=2.1 in, angle=90]{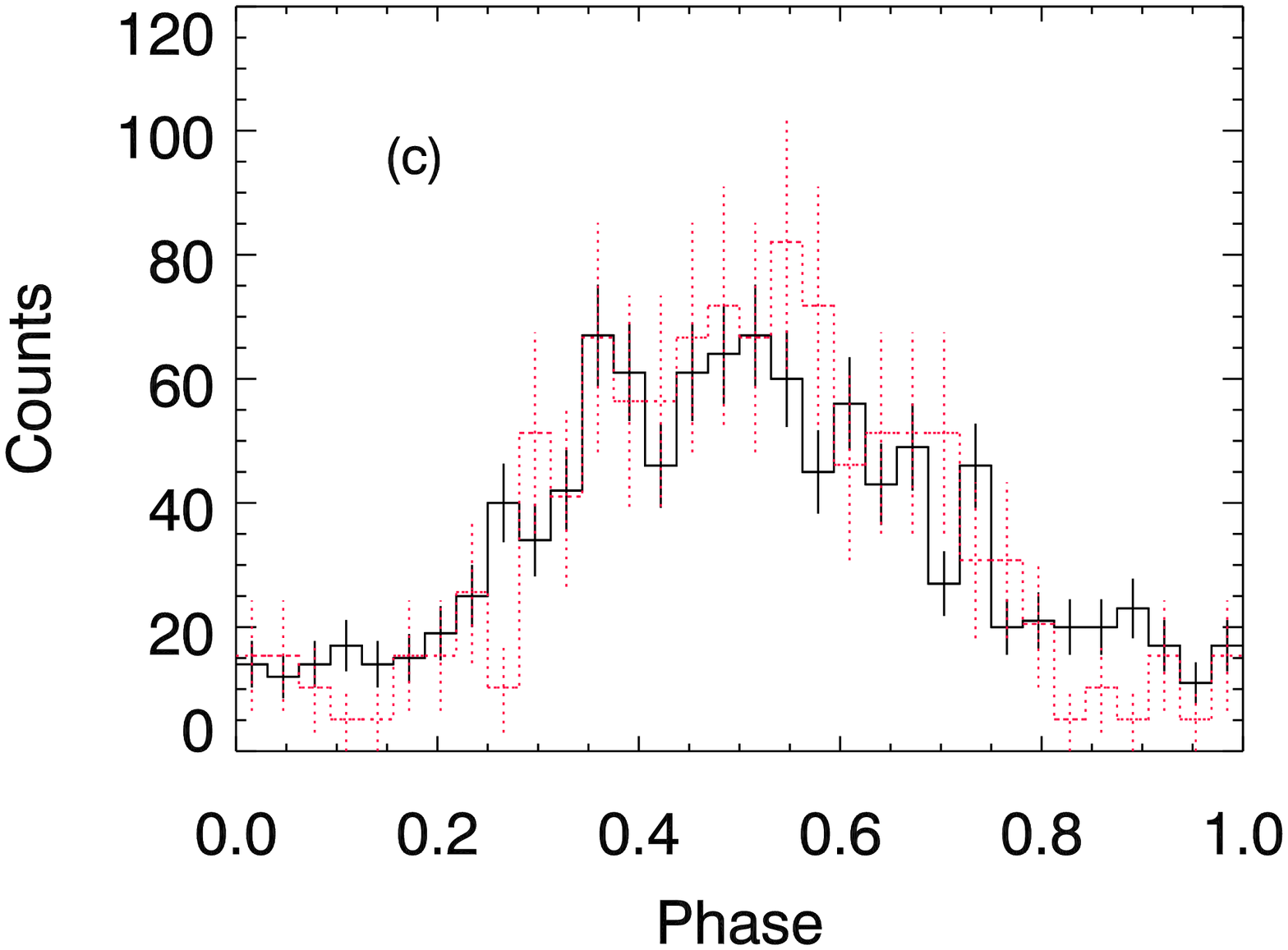} \\
\end{tabular}
\hspace{-5.0 mm}
\vspace{-5.0 mm}
\figcaption{{\em XMM-Newton} pulse profiles of pre- and post- burst time intervals of
burst 4 in the 0.5--3 keV (black) and 3--10 keV (red)
bands for the pre-burst (a), immediately post-burst in a time interval of 2--2000 s after the burst peak time
measured with {\em NuSTAR} (b),
and late post-burst (c) time intervals. Backgrounds were subtracted, and the integration times were
2 ks. The hard-band profiles were normalized to the soft-band ones for each plot.
\label{fig:xmmprofiles}
}
\end{figure*}

In order to measure the pulse period, we extracted events within radii of 60$''$ and 32$''$
for {\em NuSTAR} and {\em XMM-Newton}, respectively, and applied a barycenter correction to
the events.
We then subdivided the total observations into 51 and 48 sub-intervals so that each sub-interval
has $\sim$1200 and $\sim$2400 events to have at least $\sim$20 counts in a phase bin ($\Delta \phi=1/32$)
for {\em NuSTAR} (3--79 keV) and {\em XMM-Newton} (0.5--10 keV), respectively.
The typical duration of each sub-interval is $\sim$15 ks and $\sim$1 ks for {\em NuSTAR} and {\em XMM-Newton},
respectively, but varies depending on the source luminosity.
We then fit the pulse profile to a Gaussian plus constant function to measure the phase
of each profile. The Gaussian plus constant function describes the pulse profiles well,
and we show the measured phases in Figure~\ref{fig:residual} and examples of
pulse profiles in Figures~\ref{fig:profiles}, and \ref{fig:xmmprofiles}.
We note that cross-correlating the pulse profiles gives similar results.
The phases are fitted to a linear function $\phi(t)=\phi_0 + ft$, where $\phi_0$ is the reference phase, and
$f$ is the frequency. We did not include the frequency derivative because it is not required. In fitting, we
ignored $\sim$10--90~ks of data after the bursts because, interestingly,
there is a relatively large phase shift immediately post-burst
which contaminates the result (see Fig.~\ref{fig:residual}). From the analysis, we
found the period to be 6.46168155(6)~s, but were not able to constrain the period derivative well.

In order to see if the large phase shift during the bursts (see Fig.~\ref{fig:residual}) is related to
the rotation of the star, we
measured the shift in different energy bands ({\em NuSTAR} and {\em XMM-Newton})
and found that it is much smaller in the soft band than in the hard band
(e.g., see $T\sim$6 days in Fig.~\ref{fig:residual}).
We also verified that the energy dependence of the post-burst
phase shifts is observed in the {\em NuSTAR} and the {\em XMM-Newton} data individually,
and the large phase shift at $\sim$6 days in the {\em NuSTAR} data was seen in the
{\em XMM-Newton} hard-band data (4--10 keV) as well.
These imply that the phase shift is not due to the rotation of the source;
if it were, we would expect the shift to be independent of energy.

Figures~\ref{fig:profiles} and \ref{fig:xmmprofiles} present pre-and post-burst pulse profiles
for 1E 1048.1$-$5937 from {\it NuSTAR} and {\it XMM} data, respectively.
The pre- and post- burst pulse profiles are qualitatively similar except for the phase shifts discussed
above. Some `spiky' features appeared in some post-burst profiles
(e.g., Fig~\ref{fig:profiles}), which are likely to be
statistical fluctuations.
However, we note that another hard peak at phase $\sim$0.3 in the 3--10 keV band seemed to appear in the
{\em XMM-Newton} pulse profile immediately post-burst (see Fig.~\ref{fig:xmmprofiles}b).

\subsection{Burst Spectroscopy}
\label{burstspec}
Next, we focus on the spectra of the bursts and their tails.
The spectrum of the full data will be presented elsewhere (Archibald R.~F. 2014, in prep.).
In order to characterize a
burst spectrum, we calculated the $T_{\rm 90}$ for each event; these are shown in Table~\ref{ta:bursts}.
We note that deadtime is likely to have affected the burst light curves (Section~\ref{burstmorphology})
since we used unbinned events and did not correct for the deadtime for individual events.
A spectrum, integrated over a time interval, is less affected
since the deadtime effect is corrected for every 1-s time bin,
which is precise enough unless the spectral shape rapidly changes within the 1-s time bin.
We assume that the spectral shape (i.e., $kT$) of the source
did not change significantly in one second, and hence that the
deadtime effect is properly corrected.
\begin{figure*}
\centering
\begin{tabular}{cc}
\hspace{-13.00 mm}
\vspace{-8.00 mm}
\includegraphics[width=4.1 in]{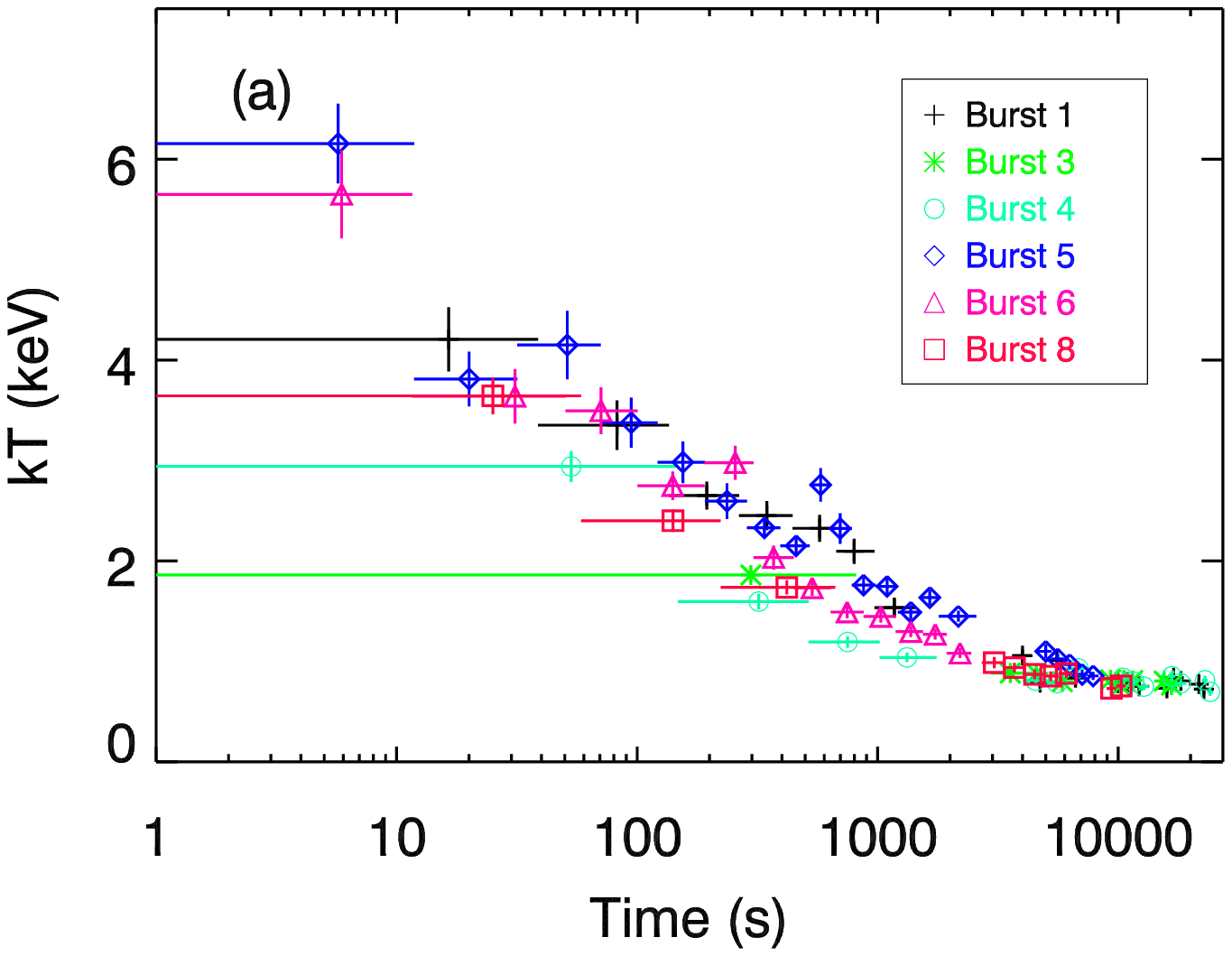} &
\hspace{-16.00 mm}
\includegraphics[width=4.1 in]{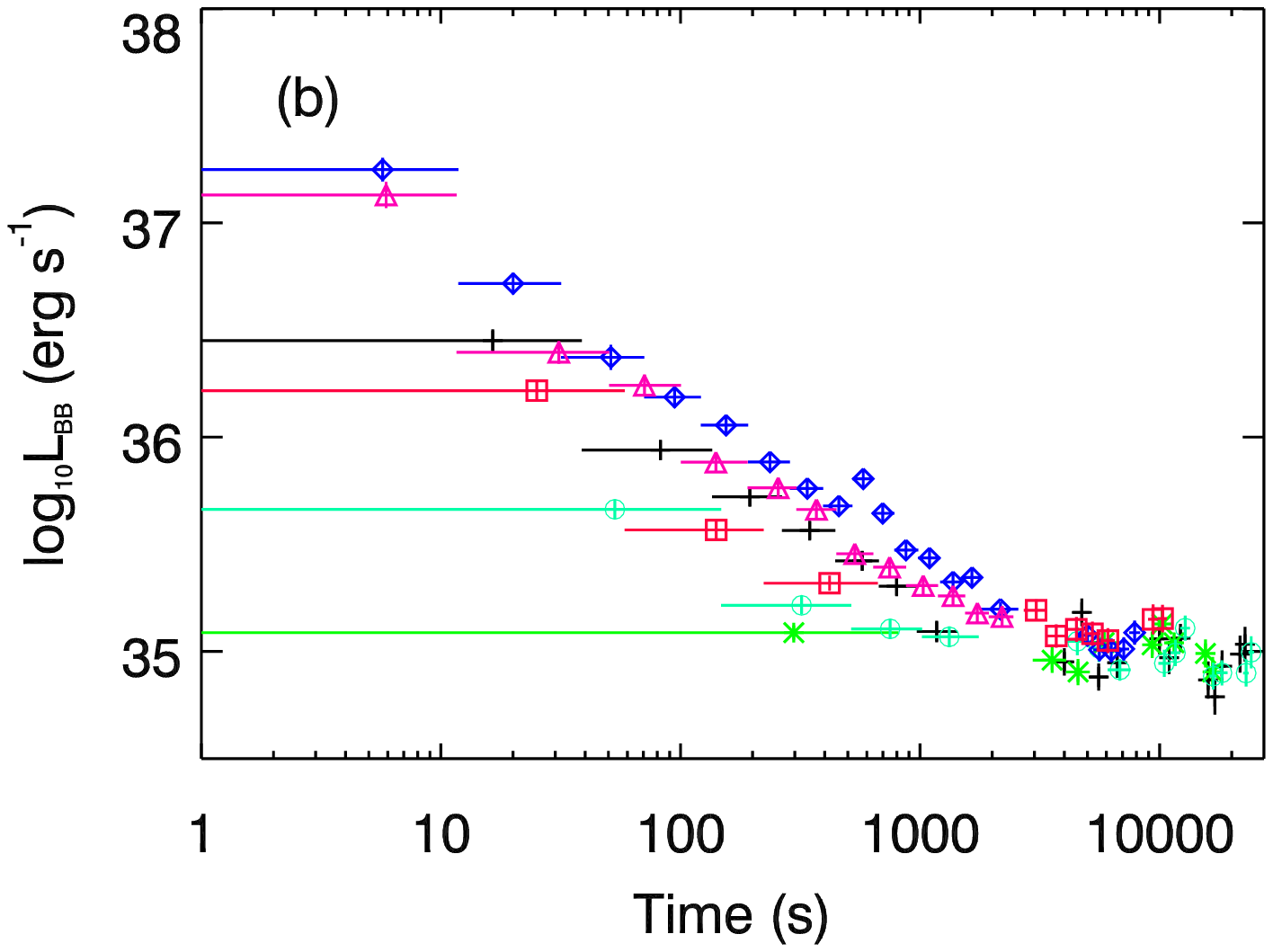} \\
\hspace{-13.00 mm}
\vspace{-12.00 mm}
\includegraphics[width=4.1 in]{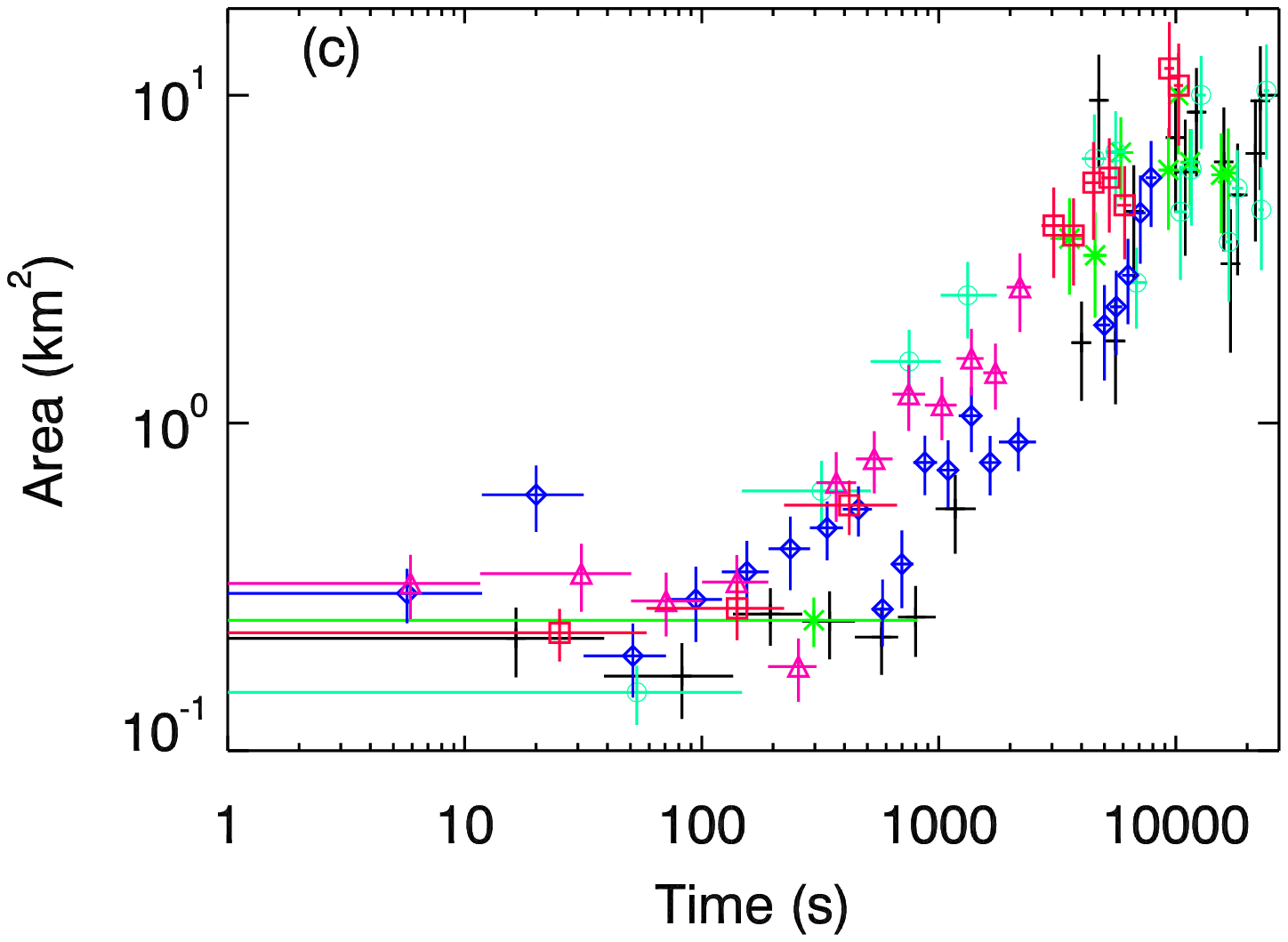} &
\hspace{-16.00 mm}
\includegraphics[width=4.1 in]{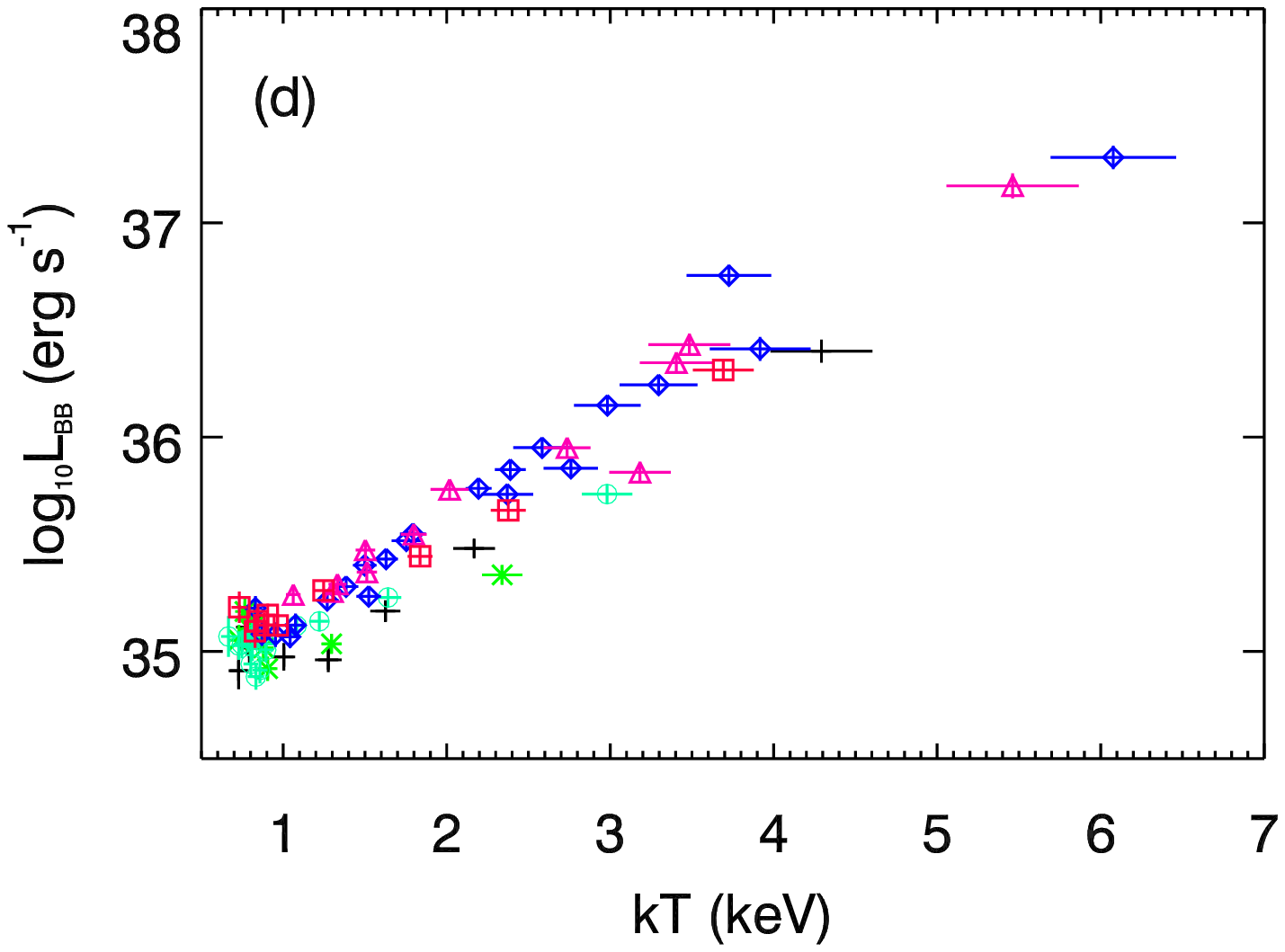} \\
\end{tabular}
\vspace{7.0 mm}
\figcaption{Time evolutions of the spectral parameters of the blackbody model for the
persistent plus tail spectra and $kT$ v.s $L_{\rm BB}$ for tail spectra of the six bursts (bursts 1, 3, 4, 5, 6, and 8).
({\it a}): blackbody temperature, $kT$, ({\it b}): bolometric luminosity, $L_{\rm BB}$, ({\it c}): blackbody area,
and ({\it d}): $kT$ v.s $L_{\rm BB}$.
\label{fig:bbradfit}
}
\end{figure*}

We extracted source events in a circle with radius 60$''$ in the time interval
$T_{\rm 90}$ for each burst. Backgrounds were extracted
from the pre-burst time intervals shown in Figure~\ref{fig:lcall} (red hatched regions).
The pre-burst spectra include photons up to $\sim$15 keV above the background and are
well described with a power law plus blackbody or two blackbody models, both having
3--79 keV flux of $4\times 10^{-12}\ \rm erg\ s^{-1}\ cm^{-2}$. The luminosity
for the assumed two blackbody model is $1\times10^{35}\ \rm erg\ s^{-1}$
(see Archibald R.~F. 2014, in prep. for more detail).
Note that the extraction times are very short,
so background contamination is very small ($\sim 0.2$ cps including the persistent source emission).

We fit the spectrum to an absorbed blackbody and an absorbed power law using the
$C$ statistic ({\ttfamily cstat} in {\ttfamily XSPEC} 12.8.1) in the 3--$E_{\rm max}$ keV band,
where $E_{\rm max}$ was between 30 and 50 keV depending on the source flux.
We verified that the fit results did not change over a broad range of the upper energy limit.
There were not many events in each spectrum, and we found that both spectral models
provided good fits. Since {\em NuSTAR} is not sensitive in the energy band below 3 keV, we were
not able to measure $N_{\rm H}$, and so we froze it at the value measured
previously \citep[$0.97\times 10^{22}\ \rm cm^{-2}$,][]{tgd+08}.

We show the best-fit parameters for the blackbody model in Table~\ref{ta:bursts}.
Note that the numbers of events in $T_{\rm 90}$
for the bursts 2, 3, 4, and 7 were smaller than $\sim$10, and the spectral parameters were not reasonably
constrained. We also tried to fit the data using the usual $\chi^2$ method
with Churazov weighting \citep[][]{cgf+96}
after grouping the spectra to have at least 5 counts per bin
found that the results are consistent with those obtained using the $C$ statistic.

Although single component models provide good fits, we also tried to fit the spectra to double blackbody models,
since magnetars sometimes show double blackbody spectra during bursts \citep[e.g.,][]{lgb+12}.
The second blackbody component
is not statistically required for any bursts in our data. However, it is possible that there was an
undetected high-temperature blackbody component as was seen by, e.g., \citet{lgb+12}. Therefore, we measure a
luminosity upper limit for a high-temperature blackbody component. We first froze the parameters of the
single blackbody fit at the values obtained above, added another high-temperature blackbody,
and scanned the parameter space of the high-temperature blackbody using the {\ttfamily steppar} command
in {\ttfamily XSPEC}. The luminosity upper limit increased with $kT$ as expected since {\em NuSTAR}
becomes less sensitive as $kT$ goes up. We measured the 90\% luminosity upper limit to be
0.2--8$\times10^{38}\ \rm erg\ s^{-1}$ for $kT=15\ \rm keV$, for example.

We also set an upper limit on flux of a possible lower temperature blackbody, similar to that
seen in the bursts of SGR~1900$+$14 \citep[][]{irm+08}, where the authors found a low-temperature
blackbody component with $kT$ as low as $\sim$2 keV having comparable luminosity with the high
temperature component. We followed the same procedure that we did above for the high-temperature
component but with $kT=2$ keV, and found that the luminosity upper
limits are 2--10$\times 10^{36}\ \rm erg\ s^{-1}$, always an order of magnitude smaller
than the bursts luminosities in Table~\ref{ta:bursts}.

\subsection{Tail Spectroscopy}
\label{specevol}
We also characterized the post-burst tail emission for the bursts with a significant tail
(bursts 1, 3, 4, 5, 6, and 8).
In order to minimize the effect of the burst,
we removed the $T_{\rm 90}$ intervals in this analysis.
We extracted events from a circular region of a radius 60$''$ centered on the source,
and the background from a source-free region on the same detector in a radius 90$''$ in the {\em NuSTAR} data.
Since the source spectrum may change significantly during a burst, we subdivided each burst
into several sub-intervals such that each interval has $\sim$200 counts, in order to have good
time resolution and to be statistically sensitive to the change in the spectral parameters in time.
The integration times were very short at early times after a burst, and
we could not collect enough background events. Therefore, for those sub-intervals we used a
longer background exposure (1~ks).

We fit the spectrum for each sub-interval to an absorbed blackbody and an absorbed power-law model using
$C$ statistic in {\ttfamily XSPEC}. The fit ranges were 3--$E_{\rm max}$, where $E_{\rm max}$ was
20--50 keV. Both blackbody and power-law models are acceptable, although blackbody models provide better
fits in general, giving fit statistics smaller by $\sim$10 on average ($\sim$200 dof).
We show the results for the blackbody fit in Figure~\ref{fig:bbradfit}a--c.
Note that the persistent emission is included in the spectra in this case.

We then tried to characterize the spectral evolution after removing the persistent emission.
The persistent level was extracted from the source region
but in the pre-burst time intervals shown in Figure~\ref{fig:lcall}.
The source seemed to return to the persistent level $\sim$2--3 ks after a burst,
and thus we analyzed only the first $\sim$2 ks after a burst.

We find that the spectral parameters and their evolution are similar to those of the
spectra with the persistent emission. This is expected because the burst tails dominate over
the persistent emission during the time intervals.

The bursts have very long tails ($\sim$ ks) as is often seen in magnetar bursts \citep[e.g.,][]{gdk11}.
We have measured the time scales of the decaying tails by fitting the light curve. However,
as we show above (Fig.~\ref{fig:bbradfit}), the spectrum evolves significantly
over a relatively short timescale,
and hence the measured light curves obtained using energy-integrated count rates will be different
for different telescopes because of differences in the energy responses.
In order to estimate the decay timescales in an
instrument-independent way, we directly measure the spectral evolution of the bursts.
We fit the time evolution of the spectral parameters for the tail
spectra including the persistent emission with a power-law decay,
\begin{equation}
\label{eqn:decay}
S = S_0 t^{-\alpha} + S_1,
\end{equation}
where $S$ is the spectral parameter, $S_0$ is the value of the parameter at $t=1\ \rm s$, $\alpha$ is
the decay index, and $S_1$ is a positive constant corresponding to the persistent emission.
We show the results in Table~\ref{ta:tail}. Note that we do not show the results for burst 3 because
we were not able to constrain its decay with the given data.

We note that there might be some less significant bursts in the tails. For example, bursts 2 and 7
occurred at $\sim$600 s and $\sim$200 s into the tail of burst 1 and 6, respectively. Also, there seems
to be an increase in $kT$ and $L_{\rm BB}$ at $\sim$600 s after burst 5 although we did not find any significant
burst. Undetected or less significant bursts may bias $\alpha$ to a smaller value.
Also note that we assumed that the luminosity at late times can freely vary in the fit,
and the best-fit values for
$S_{\rm 1}$ are 0.5--1.2$\times 10^{35}\ \rm erg\ s^{-1}$, sometimes smaller than the pre-burst
source luminosity. If we freeze $S_{\rm 1}$ at the pre-burst value of $1\times 10^{35}\ \rm erg\ s^{-1}$,
$\alpha$ becomes 0.9--1.0.

\begin{table}[t]
\vspace{0.0 mm}
\begin{center}
\caption{Parameters for the spectral evolution of the tails
\label{ta:tail}}
\scriptsize{
\begin{tabular}{ccccc} \hline\hline
Burst & $S_{\rm 0}$ & $\alpha_{\rm L}$ & $\alpha_{\rm kT}$ & $S_{\rm 1}$  \\
      & ($10^{35}\ \rm erg\ s^{-1}$) &           &  & ($10^{35}\ \rm erg\ s^{-1}$) \\ \hline
1     & 290(60) & $0.81(4)$ & $0.28(1)$ & 0.69(7) \\ 
4     & 90(50) & $0.8(1)$  & $0.57(5)$ & 0.84(6) \\ 
5     & 640(90) & $0.82(3)$ & $0.27(1)$ & 0.5(1) \\ 
6     & 560(120) & $0.87(5)$ & $0.29(1)$ & 0.7(2) \\ 
8     & 420(150) & $1.03(9)$ & $0.28(4)$ & 1.21(8) \\ \hline  
\end{tabular}}
\end{center}
\vspace{-1.0 mm}
\footnotesize{{\bf Notes.} 
Uncertainties are at the 1$\sigma$ confidence level.}\\
\vspace{-3.0 mm}
\end{table}

We also analyzed the {\em XMM-Newton} data to characterize precisely the soft-band (0.5--10 keV)
spectrum of the burst 4 tail which is the only burst detected simultaneously
with {\em XMM-Newton} and {\em NuSTAR}. We extracted source and background events
in circles with radius 32$''$ for an exposure of $\sim$400 s excluding the $T_{\rm 90}$ interval,
where the background region was $\sim$200$''$ north
of the source. The {\em XMM-Newton} count rates were 1.2/1.3 and 4.5 cps for MOS1/2 and PN, respectively.
Such count rates will result in spectral distortion of $\lapp$1\% and $\sim$2.5\% for MOS1/2 and PN, much smaller
than the statistical uncertainties we obtain below. Hence, we used all the MOS1/2 and PN data.
We grouped the spectrum to have at least 20 events per spectral bin.
We then extracted {\em NuSTAR} events for the same time interval, and binned the spectrum to have 20 events per bin,
and jointly fit the {\em XMM-Newton} and the {\em NuSTAR} data with holding $N_{\rm H}$ fixed at
$0.97\times 10^{22}\ \rm cm^{-2}$.
Note that the persistent emission was not removed in these spectra.

Single component models
were not acceptable with $\chi^2$/dof of 583/102 and 122/102 for a blackbody and a power law, respectively.
A double blackbody model was acceptable with $\chi^2$/dof=127/100.
But a blackbody plus power-law model provided a better fit ($\chi^2$/dof=98/100),
and adding one more blackbody
slightly improved the fit ($\chi^2$/dof=96/98) although it was required only with low significance
(F-test probability 40\%). Nevertheless, the parameters for the soft spectral component
($kT_{\rm l}=0.6\pm0.2$ keV and $\Gamma=2.5\pm0.4$) 
of the power law plus two blackbody model were similar to those of
the quiescent spectrum of the source \citep[][]{tgd+08},
and the hard component ($kT_{\rm h}=3.2\pm0.4$ keV) 
is similar to that of the spectrum with the persistent emission removed (see below).
When we let $N_{\rm H}$ vary, we obtain similar results as the above
with $N_{\rm H}=1.2(4)\times 10^{22}\ \rm cm^{-2}$.

We also tried to characterize the persistent-emission-removed spectrum of the combined data.
Here, the persistent level was extracted from the pre-burst time intervals in the source region
for both {\em NuSTAR} and {\em XMM-Newton}.
A blackbody model was able to fit the data well ($kT=3.0\pm0.2$ keV, $\chi^2$/dof=96/104) but a
power-law model was not as good ($\chi^2$/dof=128/104).

\subsection{Spectral Feature}
\label{specfeature}

\newcommand{\marka}{\tablenotemark{a}}
\newcommand{\markb}{\tablenotemark{b}}
\newcommand{\markc}{\tablenotemark{c}}
\newcommand{\markd}{\tablenotemark{d}}
\begin{table*}[t]
\vspace{0.0 mm}
\begin{center}
\caption{Best-fit spectral parameters for the spectrum of the first 150 s after $T_0$ for burst 5
\label{ta:spec}}
\scriptsize{
\begin{tabular}{cccccccc} \hline\hline
Model\marka & $N_{\rm H}$\markb & $kT/\Gamma$ & $L_{\rm BB}/F_{\rm PL}$\markc & $E_{\rm G}$\markd & $\sigma_{\rm G}$ & $N_{\rm G}$ & $\chi^2$/dof \\ 
	   & ($10^{22}\ \rm cm^{-2}$)	& (keV/ ) &  & (keV) & (keV) & (ph $\rm cm^{-2}\ \rm s^{-1}$) & \\ \hline
BB         & 0.97 & 4.1(2) & 30(2) & $\cdots$ & $\cdots$ & $\cdots$ & 70/42 \\
BB + Gauss & 0.97 & 3.8(2) & 25(2) & 13.2(3)  & 1.3(3)   & $2.1(5)\times 10^{-3}$ & 39/39 \\
PL         & 0.97 & 0.83(7) & 12(1) & $\cdots$ & $\cdots$ & $\cdots$ & 108/42 \\ 
PL + Gauss & 0.97 & 1.0(1) & 0.8(1) & 12.7(2) & 1.7(3) & $3.4(6)\times 10^{-3}$ & 44/39 \\ \hline
\end{tabular}}
\end{center}
\vspace{-1 mm}
\footnotesize{{\bf Notes. Uncertainties are at the $1\sigma$ confidence level.}}\\
$^{\rm a}${ BB: blackbody, PL: power law, Gauss: Gaussian line profile.}\\
$^{\rm b}${ Frozen.}\\
$^{\rm c}${ Bolometric luminosity in units of $10^{35}\ \rm erg\ s^{-1}$ for the blackbody model, 
and the absorption corrected 3--79 keV flux in units of $10^{-9}\ \rm erg\ s^{-1} cm^{-2}$ for the power-law model.}\\ 
$^{\rm d}${ Subscript G is used for the Gaussian parameters: line energy ($E_{\rm G}$), line width ($\sigma_{\rm G}$), and
normalization ($N_{\rm G}$).}\\
\vspace{1.0 mm}
\end{table*}

We searched for the spectral feature at $\sim$13 keV previously observed from
this source's burst spectra \citep[][]{gkw02,dkg09,gdk11}.
As the feature was observed in the $\lapp$2 sec spectra in the past, we
checked if a line feature was statistically required in the burst spectra
(see Section~\ref{burstspec}) but found that it is not.
However, as we increased the integration time,
we started seeing an enhancement in counts at $\sim$13 keV for burst 5.

\begin{figure}
\centering
\vspace{-2.00 mm}
\hspace{5.00 mm}
\includegraphics[width=2.7 in, angle=90, origin=br]{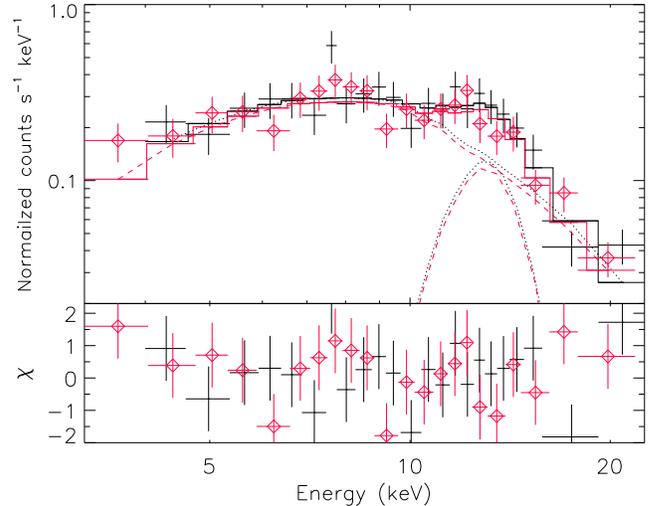} 
\figcaption{A blackbody plus Gaussian line fit for the 150-s {\em NuSTAR} spectrum of burst 5.
\label{fig:feature}
}
\end{figure}

In order to see if a line feature is statistically required for the bursts and their
tail emission, we first extracted a spectrum from the first 150~s of the
brightest burst (burst 5).
The background was extracted from the pre-burst interval in the source region.
We then fit the spectrum in the 3--30 keV band
with continuum models (blackbody, power law, and bremsstrahlung), and found that none could
describe the data well with $\chi^2$/dof of 70/42 (blackbody), 108/42 (power law),
and 147/42 (bremsstrahlung).
We also tried combinations of the continuum models, but none improve the fit significantly with
$\chi^2$/dof being 67/40, 108/40, 67/40, 67/40 for a double blackbody, a double power-law,
a blackbody plus power-law, and a blackbody plus bremsstrahlung models, respectively.
In particular, in all these two-component trials, the best-fit parameters
of the second component are trivial; the amplitude is
not constrained at all, and $kT$ or $\Gamma$ become ridiculously small or large.
However, adding a Gaussian line to a blackbody model provided an acceptable fit ($\chi^2$/dof = 38/40). 
We show the spectrum and the blackbody plus Gaussian fit in Figure~\ref{fig:feature},
and the results are presented in Table~\ref{ta:spec}.
We repeated the same procedure for spectra in different time
intervals (e.g., 100~s, 200~s, and 300~s), and found that the fit improves significantly
($\Delta \chi^2$$\gapp 20$) by adding a Gaussian line to the blackbody model, from which we draw
the same conclusion as we did with the 150-s spectrum.
We conducted the same analysis with the other bursts and found that
a blackbody model alone was able to describe their spectra well (for example,
$\chi^2$/dof=39/35, $p\sim0.1$ for burst 6).

We used simulations in order to calculate the true significance of the feature in burst 5 and its tail.
The spectrum of the source following a burst evolved on a time scale of 150-s.
Therefore, we conducted simulations with temporally
evolving blackbody spectra. We first divided the 150-s spectrum of burst 5
into five time intervals as shown in Figure~\ref{fig:bbradfit} and the $T_{\rm 90}$ interval,
and fit the six spectra to blackbody models after removing the energy range of
the feature (11.5--14 keV). We then simulated
the six spectra using {\ttfamily fakeit} in {\ttfamily XSPEC} with the response
and background for the actual data, and merged them into one spectrum that represents a 150-s spectrum.
For each simulation, we first fit the spectrum to a blackbody,
and then calculated the improvement of the fit ($\Delta \chi^2$) by adding a Gaussian line.
For the latter fit, we scanned the energy range of the spectrum (4--22 keV) with a step of 1 keV
for the initial value of the line energy because the narrow feature can make the fit fall in
a local minimum around the initial value. We counted the number of occurrences in which the
improvement of the fit in a simulation was larger than that seen in the actual
data fit ($\Delta \chi^2=31$), and found that this did not occur in 10,000 simulations.
The significance of the feature would be even higher if we scan only a smaller
energy range (e.g., 11--15 keV) for the Gaussian line energy guided by
the previous {\em RXTE} measurements.

We found that a blackbody model fits the spectra of the feature-removed
150-s data as well as the above simulations.
Therefore, we also conducted simulations with a single blackbody
continuum model, and found that improvement of fit measured by $\Delta \chi^2$
by adding a Gaussian line model was always smaller than 31 in 10,000 simulations.

We investigated if the feature exists in the spectra without the burst.
We removed the $T_{\rm 90}$ interval from burst 5 and
extracted tail spectra for 100 s, 150 s, 200 s, and 300 s exposures.
We then applied the same fitting procedure and found that the same conclusion is valid;
for example, single or combinations of continuum models do not
fit the 150-s spectrum well with $\chi^2$/dof of 75/39 (blackbody), and 164/39 (power law)
while adding a Gaussian line to a blackbody improves the fit significantly, making it
acceptable ($\chi^2$/dof = 42/36). The best-fit Gaussian line parameters are statistically
consistent with those we obtained above.

We checked if the spectral feature shifts in energy with phase as seen in the magnetar
SGR~0418$+$5729 \citep[][]{tem+13}. We produced a 2-D energy versus phase image as shown
in \citet[][]{tem+13}, and found no evidence of a shift. This may be due to the paucity
of counts in our case. We also tried to see if the feature is more prominent in some phase
intervals than the others, and could not draw any statistically significant conclusion. 

\medskip
\section{Discussion}
\label{sec:disc}
We have reported the first detection of X-ray bursts from magnetar 1E 1048.1$-$5937
with a focusing hard X-ray telescope.
Table~\ref{ta:bursts} summarizes the temporal and spectral properties of the
eight bursts detected with {\em NuSTAR}.
The short timescale behaviors of the bursts are different from burst to burst,
having $T_{\rm r}$  and $T_{\rm f}$
of milliseconds to seconds. All the bursts have relatively long tails (see Fig.~\ref{fig:bbradfit})
and their spectra evolve with time. We also found that their peak times are random in phase.

\subsection{Temporal Properties of the Bursts and their Tails}
Magnetar bursts, including those we have observed for 1E~1048.1$-$5937, generally
exhibit a fast rise and fall,
which has been suggested to result from crustal fracture \citep[][]{tlk02}
and/or magnetic reconnection \cite[][]{l03}. 
The pulse phase of the burst peak was previously used to reconstruct the event location
and constrain the emission geometry \citep[][]{wkg+05}. Interestingly, the bursts appear
to occur at random pulse phases (Table~\ref{ta:bursts}), and bursts 2 and 5 may be emitted
far from the magnetic axis of the star. Note, however, that the location reconstruction
assumes emission beaming along the magnetic field lines, which may not be realistic.
A quasi-isotropic burst could be visible from any direction unless it is eclipsed
by the neutron star. The probability of eclipse is significantly reduced by gravitational
light bending, which makes $\approx 3/4$ of the star visible to a distant observer \citep[][]{b02}.

A burst can have tail emission which is produced by
the residual heat of the crustal fracturing \citep[][]{let02} or
bombardment of the stellar surface by the magnetospheric particles \citep[][]{b09}.
The tail emission lasts much longer than a burst, and allows us to sample many full rotations.
Studying pulse profiles during burst tails may provide insight into the tail emission region.
Interestingly, we find clear phase shifts in some post-burst profiles
(e.g., bursts 4 and 5, see Figs.~\ref{fig:residual} and \ref{fig:profiles}),
which are not caused by rotation (Section~\ref{timingana}).
We conclude that the source of tail emission, e.g. the hot spot produced by the burst,
is slightly shifted in longitude relative to the source of persistent emission.

The pulse profiles measured by XMM-Newton (Fig.~\ref{fig:xmmprofiles}) also hint at
the appearance of a new hot spot. The immediate post-burst pulse profile shows
an additional peak at phase $\sim$0.3 (Fig.~\ref{fig:xmmprofiles}b),
which may corresponds to the new hot spot. Its emission added to the pulsed
persistent source results in a phase shift, which returns to zero as the luminosity
of the new hot spot decays and the persistent source dominates again (Fig.~\ref{fig:residual}).

\subsection{Burst Spectra and Their Evolution}
Interesting correlations between burst spectral properties have been seen in some magnetars.
For example, \citet{gkw+01} reported an anti-correlation between fluence and hardness ratio in the bursts of
SGR~1806$-$20 and SGR~1900$+$14; however, \citet{gkw04} observed the opposite trend
between the same properties in the bursts of 1E~2259$+$589. While detailed studies of such correlations here are
not possible due to the limited number of bursts and small statistics in each burst,
we note that there seems to be a hint of a positive correlation between
$L_{\rm 90}$ and $kT$ (see Table~\ref{ta:bursts}).

The source flux returns to the pre-burst value on the kilo-second timescale.
Such a short timescale suggests that the crustal heating was not deep,
as the large specific heat of the deep crust would hold the energy longer
and delay the decay \citep[][]{kew+03}. The spectrum of tail emission at $\sim$10--1000 s can be fitted
by a blackbody or a power-law model; the blackbody model provides a better fit
when the persistent emission is subtracted.
The evolution of the blackbody luminosity $L_{\rm BB}$ can be described as a power law
with the decay index of 0.8--0.9 (Table~\ref{ta:tail}).\footnote{The true decay may be somewhat
steeper when the persistent emission of $10^{35}\ \rm erg\ s^{-1}$ is subtracted;
besides the measured decay index might be affected by undetected low-significance bursts
during the tail emission.} Similar flux decay indices were measured in the 2004 bursts of
1E~1048$-$5937 \citep[$0.82\pm0.05$]{gkw06} and XTE~J1810$-$197 \citep[$0.92\pm0.02$][]{wkg+05}.
The indices are not far from the crustal cooling model with $L_{\rm BB} \propto t^{-2/3}$ \citep[][]{let02},
which was previously used for later time evolution ($t\gapp10,000$ s).
The j-bundle untwisting \citep[][]{b09} is usually invoked on even longer timescales,
which are associated with the resistive evolution of the magnetosphere.

Note the increase in blackbody area (Fig.~\ref{fig:bbradfit}c) following a burst.
We argue that this is an observational effect rather than a real increase in the physical size.
If a burst produced a local hot spot as we argued above, the measured size of the blackbody
area would represent the small burst spot at early times, and the large persistent spot later.
In order to disentangle the effects of the hot spot and the persistent spot,
we also measured the size evolution with the persistent-emission-removed spectrum.
The tail was detected significantly above background only for the first $\sim$1 ks,
and we found that the blackbody area evolved in a similar manner
to the trend of the first $\sim$1 ks in Figure~\ref{fig:bbradfit}; no clear increase
in the size was observed.

In magnetar cooling scenarios, an excited magnetar cools by emitting thermal photons and/or untwisting
a $j$-bundle. In both cases, a correlation between flux and hardness is expected \citep[][]{tlk02, l03, og07}.
Such a correlation has been generally observed in magnetars' short-term and
long-term cooling \citep[e.g.,][]{wkg+05, gkw06, sk11, aka+13} with some exceptions
\citep[e.g.,][]{akt+12, kab+14}.
We investigated if such a correlation exists in the tail spectra and found
a clear correlation between $kT$ and $log_{10}L_{\rm BB}$ (see Fig.~\ref{fig:bbradfit}d),
implying that the burst tails also exhibited a correlation between flux and hardness.
Whether the origin of this
correlation is the same as for that of the long-term cooling is not clear.

In principle, tail radiation could be the burst ``echo'' produced by dust scattering around
the magnetar \citep[][]{tem+10}. In this scenario, the tail spectrum should soften with time
(a result of the scattering cross section being smaller at higher energies);
whether or not this is consistent with the observed spectral evolution is unclear.
Furthermore, the observed increases in the pulsed flux immediately after the bursts cannot
be produced by dust scattering, suggesting that the tail is emitted by the magnetar itself.

\subsection{Spectral Feature}
We find a spectral feature at $\sim$13 keV in the tail emission.
Spectral features at a similar energy were previously
observed from the burst and tail spectra of several sources, but only with
{\em RXTE} \citep[1E~1048.1$-$5937, XTE~J1810$-$197, 4U~0142$+$61][]{gkw02,wkg+05,gkw06,dkg09,gdk11}.
The line energy we found is similar to those previously reported.
We note that this is the first detection of the feature
with an instrument other than {\em RXTE}, which demonstrates the effect is not instrumental.
We note that the line flux we measured is only $\sim$10\% of that previously
reported \citep[][]{gkw02}, which could be simply
due to the long integration time we used (150-s versus 1-s).

If we interpret the line-like emission as an electron cyclotron feature,
the line energy implies that the magnetic field strength
is $\sim$10$^{12}\ \rm G$ in the emission region. A magnetic field strength of $\sim$2$\times10^{15}\ \rm G$
could be inferred if we interpret the feature due to a proton cyclotron emission,
and the line width implies a $\sim$10\% change in the magnetic field strength in the emission region.
The three detections of the feature
from 1E~1048.1$-$5937 \citep[][and this work]{gkw02,dkg09}
have similar properties (e.g., line energy and width), and probably originated from the same region of the star.
If so, it is unlikely that a physical structure can be sustaining at a height of $\sim$70 km from
the stellar surface (where $B$$\sim$$10^{12}\ \rm G$)
to power bursts and line features multiple times over a decade.
However, a strong multipolar magnetic field ($10^{15}\ \rm G$) near
the surface in a volume of $\sim$1 km$^3$ has enough energy to power multiple bursts
and the line feature,
suggesting that the line feature could be from proton cyclotron emission.
Nevertheless, an interesting question is how different sources
which may have different magnetic fields
(1E~1048.1$-$5937, XTE~J1810$-$197, and 4U~0142$+$61) with different spectral
and temporal properties show similar features  at such similar energies.

\section{Conclusions}
\label{sec:concl}
We presented detailed spectral and temporal analyses of eight
X-ray bursts from magnetar 1E 1048.1$-$5937 detected with {\em NuSTAR},
one of which was simultaneously observed with {\em XMM-Newton}.
The bursts exhibited a fast rise and decay with $T_{\rm 90}$ intervals of 1--4 s, and
their spectra can be described with single blackbody models having $kT\sim$ 6--8~keV for
the $T_{\rm 90}$ intervals. All the bursts showed tail emission which can be
described with temporally relaxing blackbody models.
The flux relaxations of the tail emission followed a power-law decay having decay indices 0.8--0.9.
We confirm the existence of an emission feature at $\sim$13 keV observed
in burst and tail spectra of the source in the past.
Finally, we note that similar spectral features at a similar energy have been seen in bursts
of several magnetars with different spectral and temporal properties. This requires further
theoretical interpretations.

\medskip

This work was supported under NASA Contract No. NNG08FD60C, and  made use of data from the {\it NuSTAR} mission,
a project led by  the California Institute of Technology, managed by the Jet Propulsion  Laboratory,
and funded by the National Aeronautics and Space  Administration. We thank the {\it NuSTAR} Operations,
Software and  Calibration teams for support with the execution and analysis of  these observations.
This research has made use of the {\it NuSTAR}  Data Analysis Software (NuSTARDAS) jointly developed by
the ASI  Science Data Center (ASDC, Italy) and the California Institute of  Technology (USA). V.M.K. acknowledges support
from an NSERC Discovery Grant, the FQRNT Centre de Recherche Astrophysique du Qu\'ebec,
an R. Howard Webster Foundation Fellowship from the Canadian Institute for Advanced
Research (CIFAR), the Canada Research Chairs Program and the Lorne Trottier Chair
in Astrophysics and Cosmology.
A.M.B. acknowledges the support by NASA grants NNX10AI72G and NNX13AI34G.


\begin{thebibliography}{48}
\expandafter\ifx\csname natexlab\endcsname\relax\def\natexlab#1{#1}\fi

\bibitem[{{An} {et~al.}(2012){An}, {Kaspi}, {Archibald}\& {Tomsick}}]{akt+12}
{An}, H., {Kaspi}, V.~K., {Tomsick}, J.~A., {Cumming}, A., {Bodaghee}, A.,
{Gotthelf}, E.~V., \& {Rahoui}, F. 2012, \apj, 757, 68

\bibitem[{{An} {et~al.}(2013){An}, {Kaspi}, {Archibald}\& {Cumming}}]{aka+13}
{An}, H., {Kaspi}, V.~K., {Archibald}, R., \& {Cumming}, A. 2013, \apj, 763, 82

\bibitem[{{Beloborodov}(2002)}]{b02}{Beloborodov}, A.~M. 2002, \apj, 566, L85 

\bibitem[{{Beloborodov}(2009)}]{b09}{Beloborodov}, A.~M. 2009, \apj, 703, 1044,

\bibitem[{{Churazov} {et~al.}(1996){Churazov}, {Gilfanov}, {Forman}, \&
{Jones}}]{cgf+96}
{Churazov}, E., {Gilfanov}, M., {Forman}, W., \& {Jones}, C. \apj, 471, 673

\bibitem[{Dib} {et~al.}(2009){Dib}, {Kaspi}, \& {Gavriil}]{dkg09}
{Dib}, R., {Kaspi}, V.~M., \& {Gavriil}, F.~P. 2009, \apj, 702, 614

\bibitem[{{Duncan} \& {Thompson}(1992)}]{dt92}
{Duncan}, R.~C., \& {Thompson}, C. 1992, \apj, 392, L9

\bibitem[{Durant} \& {van Kerkwijk}(2006){Durant} \& {van Kerkwijk}]{dv06}
{Durant}, M., \& {van Kerkwijk}, M.~H. 2006, \apj, 650, 1070

\bibitem[Feroci et al.(2001)]{fhd+01}
Feroci, M., Hurley, K.,  Duncan, R.~C., \& Thompson, C.\ 2001, \apj, 549, 1021

\bibitem[{Gavriil}, {Dib}, \& {Kaspi}(2011){Gavriil}, {Dib}, \& {Kaspi}]{gdk11}
{Gavriil}, F.~P., {Dib}, R., \& {Kaspi}, V.~M. 2011, \apj, 736, 138

\bibitem[{Gavriil} {et~al.}(2002){Gavriil}, {Kaspi}, {Woods}]{gkw02}
{Gavriil}, F.~P., {Kaspi}, V.~M. and {Woods}, P.~M., 2002, \nat, 419, 142

\bibitem[{Gavriil} {et~al.}(2004){Gavriil}, {Kaspi}, {Woods}]{gkw04}
{Gavriil}, F.~P., {Kaspi}, V.~M. and {Woods}, P.~M., 2004, \apj, 607, 959

\bibitem[{Gavriil} {et~al.}(2006){Gavriil}, {Kaspi}, {Woods}]{gkw06}
{Gavriil}, F.~P., {Kaspi}, V.~M. and {Woods}, P.~M., 2006, \apj, 641, 418

\bibitem[{G{\"o}{\v g}{\"u}{\c s}} {et~al.}(2006){G{\"o}{\v g}{\"u}{\c s}},
{Kouveliotou}, {Woods}, {Thompson}, {Duncan}, {Briggs}]{gkw+01}
{G{\"o}{\v g}{\"u}{\c s}}, E., {Kouveliotou}, C., {Woods}, P.~M.,
{Thompson}, C., {Duncan}, R.~C., \& {Briggs}, M.~S. 2001, \apj, 558, 228

\bibitem[G{\"o}{\v g}{\"u}{\c s} et al.(2011)]{gwk+11}
G{\"o}{\v g}{\"u}{\c s}, E., Woods, P.~M., Kouveliotou, C., et al.\ 2011,  \apj, 740, 55

\bibitem[{{Harrison} {et~al.}(2013){Harrison}}]{hcc+13}
{Harrison}, F.~A., {Craig}, W.~W., {Christensen}, F.~E. et~al. 2013, \apj, 770, 103

\bibitem[{{Israel} {et~al.}(2008){Israel}}]{irm+08}
{Israel}, G.~L., {Romano}, P., {Mangano}, V. et~al. 2008, \apj, 685, 1114

\bibitem[{Kaspi} {et~al.}(2003){Kaspi}, {et~al.}]{kgw+03}
{Kaspi}, V.~M., {Gavriil}, F.~P., {Woods}, P.~M., {Jensen}, J.~B.,
{Roberts}, M.~S.~E., {Chakrabarty}, D. 2003, \apj, 588L, 93

\bibitem[{Kaspi} {et~al.}(2014){Kaspi}, {et~al.}]{kab+14}
{Kaspi}, V.~M., {Archibald}, R.~F., {Bhalerao}, V. et~al. 2014, \apj, 786, 84

\bibitem[{Kouveliotou} {et~al.}(1998){Dieters}, {Strohmayer}, {van Paradijs},
{Fishman}, {Meegan}, {Hurley}, {Kommers}, {Smith}, {Frail}, \& {Murakami}]{kds+98}
{Kouveliotou}, C., {Dieters}, S., {Strohmayer}, T. et~al. 1998, \nat, 393, 235

\bibitem[{Kouveliotou} {et~al.}(2003){Kouveliotou}, {Eichler}, {Woods}, \& {et~al.}]{kew+03}
{Kouveliotou}, C., {Eichler}, D., {Woods}, P.~M. \& et~al. 2003, \apj, 596, L79

\bibitem[Lenters et al.(2003)]{lwg+03}
Lenters, G.~T., Woods,  P.~M., Goupell, J.~E. et al.\ 2003, \apj, 587, 761
%
\bibitem[{Lin} {et~al.}(2012){Lin}, {G{\"o}{\v g}{\"u}{\c s}}, {Baring}, 
{Granot}, {Kouveliotou}, {Kaneko}, {van der Horst},
{Gruber}, {von Kienlin}, {Younes}, {Watts},
{Gehrels}, N.]{lgb+12}
{Lin}, L., {G{\"o}{\v g}{\"u}{\c s}}, E., {Baring}, M.~G. et~al. 2012, \apj, 756, 54

\bibitem[{Lyubarsky} {et~al.}(2002){Lyubarsky}, {Eichler} \& {Thompson}]{let02}
{Lyubarsky}, Y., {Eichler}, D., \& {Thompson}, C. 2002, \apj, 580, L69

\bibitem[{Lyutikov}(2003){Lyutikov}]{l03}
{Lyutikov}, M. 2003, \mnras, 346, 540

\bibitem[see Olausen \& Kaspi(2014)]{ok14}
{Olausen}, S.~A., \& {Kaspi}, V.~M. 2014, \apjs, 212, 6

\bibitem[{\"O}zel \& G{\"u}ver(2007)]{og07}
{\"O}zel, F., \& G{\"u}ver, T. 2007, \apjl, 659, 141

\bibitem[{Scholz} {et~al.}(2011){Scholz}, {Kaspi}]{sk11}
{Scholz} P., \& {Kaspi}, V.~M. 2011, \apj, 739, 94

\bibitem[{Tam} {et~al.}(2008){{Tam}, {Gavriil}, {Dib}, {Kaspi}, {Woods}, \& {Bassa}}]{tgd+08}
{Tam}, C.~R., {Gavriil}, F.~P., {Dib}, R. et~al. 2008, \apj, 677, 503

\bibitem[{Thompson} \& {Duncan}(1995)]{td95}
{Thompson}, C., \& {Duncan}, R.~C., 1995, \mnras, 275, 255

\bibitem[{Thompson} \& {Duncan}(1996)]{td96}
{Thompson}, C., \& {Duncan}, R.~C., 1996, \apj, 473, 322

\bibitem[{Thompson {et~al.}(2002)Thompson, Lyutikov, \& Kulkarni}]{tlk02}
Thompson, C., Lyutikov, M., \& Kulkarni, S.~R. 2002, \apj, 574, 332

\bibitem[{Tiengo} {et~al.}(2010){{Tiengo}, {Esposito}, {Mereghetti}, }]{tem+10}
{Tiengo}, A., {Esposito}, P., {Mereghetti}, S. et~al. 2010, \apj, 710, 227

\bibitem[{Tiengo} {et~al.}(2013){{Tiengo}, {Esposito}, {Mereghetti}, {Turolla}, 
{Nobili}, {Gastaldello}, {G{\"o}tz}, {Israel},
{Rea}, {Stella}, {Zane}, {Bignami}}]{tem+13}
{Tiengo}, A., {Esposito}, P., {Mereghetti}, S. et~al. 2013, \nat, 500, 312

\bibitem[{van der Horst} {et~al.}(2012)]{vkg+12}
{van der Horst}, A.~J., {Kouveliotou}, C., {Gorgone}, N.~M. et~al. 2012, \apj, 749, 122

\bibitem[{Vasisht} \& {Gotthelf}(1997){Vasisht}, {Gotthelf}]{vg97}
{Vasisht}, G., \& {Gotthelf}, E.~V. 1997, \apj, 486, L129

\bibitem[{Woods} {et~al.}(2005){{Woods}, {Kouveliotou}, {Gavriil}, {Kaspi}, {Roberts},
{Ibrahim}, {Markwardt}, {Swank}, {Finger}}]{wkg+05}
{Woods}, P.~M., {Kouveliotou}, C., {Gavriil}, F.~P. et~al. 2005, \apj, 629, 985
\end{thebibliography}
\end{document}